\documentclass[twocolumn,showpacs,preprintnumbers,amsmath,amssymb,ctexart]{revtex4}
\usepackage{pifont}
\usepackage{CJK}
\usepackage{tipa}
\usepackage{graphicx}
\usepackage{dcolumn}
\usepackage{slashbox}
\usepackage{bm}
\usepackage{multirow}

\begin{document}
\title{Hadronic production of $D(2550)$, $D^*(2600)$, $D(2750)$, $D^*_1(2760)$ and $D^*_3(2760)$}

\author{Ze Zhao, Yu Tian and Ailin Zhang
\footnote{Corresponding author: zhangal@staff.shu.edu.cn}}
\affiliation{Department of Physics, Shanghai University, Shanghai 200444, China}
\begin{abstract}
Hadronic decays of radially excited $2D$ $D(2^3D_1)$, $D(2^3D_3)$, $D(2D_2)$ and $D(2D^\prime_2)$ have been studied in a $^3P_0$ model. All OZI-allowed decay channels of these $2D$ $D$ resonances have been given, and relevant decay widths have been calculated. $D(2550)$, $D^*(2600)$, $D(2750)$, $D^*_1(2760)$ and $D^*_3(2760)$ can be produced in hadronic decays of $D(2^3D_1)$, $D(2^3D_3)$, $D(2D_2)$ and $D(2D^\prime_2)$. In different assignments, hadronic decay widths and some relevant ratios from the $2D$ $D$ resonances to $D(2550)$, $D^*(2600)$, $D(2750)$, $D^*_1(2760)$ or $D^*_3(2760)$ final states have been predicted, which may provide some more information to identify these resonances in forthcoming experiments.
\end{abstract}
\pacs{13.25.Ft; 12.39.-x   \\
Keywords: Hadronic decay, $^3P_0$ model, $2D$ $D$ resonances}

\maketitle

\section{INTRODUCTION}
It is believed that S-wave and P-wave $D$ and $D_s$ charmed mesons without radial excitation have well established. $D$ and $D^*$ are the two S-wave $D$ mesons with $J^P=0^-,1^-$, $D^*_0(2400)$, $D_1(2420)$, $D^\prime_1(2430)$ and $D^*_2(2460)$ are the four P-wave $D$ mesons with $J^P=0^+,1^+,1^+,2^+$, respectively. $D_s$ and $D^*_s$ are the two S-wave $D_s$ mesons with $J^P=0^-,1^-$, $D^*_{s0}(2317)$, $D_{s1}(2460)$, $D_{s1}(2536)$ and $D^*_{s2}(2573)$ are the four P-wave $D_s$ mesons with $J^P=0^+,1^+,1^+,2^+$, respectively. In recent years, more and more higher excited $D$ and $D_s$ have been observed by B factories and LHCb~\cite{pdg}, which raises much interest on these new observed charmed resonances. It is a challenge to to identify them and to find more new charmed resonances.

\par $D(2550)$, $D^*(2600)$, $D(2750)$ and $D^*(2760)$ are first observed in inclusive $e^+e^-$ collisions by BaBar collaboration~\cite{Benitez}
$$e^+e^-\to c\bar c\to D^{(\star)}\pi X,$$
where $X$ is other system. It is found that $D(2550)$ and $D^*(2600)$ are consistent with the predicted radial excitations $2S$ $^1S_0$ and $^3S_1$, respectively, through the mass values and $\cos\theta_H$ distributions in this experiment.

In $pp$ collision~\cite{lhcb1}, two natural parity ($J^P=0^+,1^-,2^+,\cdots$ with $P=(-1)^J$) resonances $D^\star_J(2650)^0$ and $D^\star_J(2760)^0$ are observed in $D^{\star+}\pi^-$ channel from the mass spectrum and the angular distributions. These two resonances are believed the $D^*(2600)$ and $D^*(2760)$ observed in $e^+e^-$ collisions, respectively. Two unnatural parity ($J^P=0^-,1^+,2^-,\cdots$) states $D_J(2580)^0$ and $D_J(2740)^0$ are also observed in $D^{\star+}\pi^-$ channel. These two resonances are believed the $D(2550)$ and $D(2750)$. In inclusive decays, the spin and parity are difficult to determine.

In addition to inclusive production in $e^+e^-$ and $pp$ collisions, highly excited heavy flavor resonances are also produced in exclusive $B$ decays. In decay $B^-\to D^+K^-\pi^-$~\cite{lhcb2}, $D^\star_1(2760)$ is observed, and in decay $B^0\to \bar D^0\pi^+\pi^-$~\cite{lhcb3}, $D^\star_3(2760)$ is observed. In these decays, the spin-parity is determined through a Dalitz plot technique.

In charmed-strange mesons sector, $D^*_{sJ}(2860)$ produced in $e^+e^-$ and $pp$ collisions by BaBar and LHCb, respectively, is found to consist of $D^*_{s1}(2860)$ and $D^*_{s3}(2860)$~\cite{lhcb4,lhcb5}. Similarly, $D^*(2760)$ produced in $e^+e^-$ and $pp$ collisions is also found to consists of $D^\star_1(2760)$ and $D^\star_3(2760)$~\cite{lhcb2,lhcb3}. Accordingly, $D^*(2760)$ is not analyzed in our study.

Hadronic decays of heavy-light mesons exhibit much information of the quark dynamics of hadrons, and have been extensively studied for a long time~\cite{flavor,gk,ehq,pe,st,fs,zhao,colangelo,gm}. Since the observation of $D(2550)$, $D^*(2600)$, $D(2750)$ and $D^*(2760)$, hadronic decays of these resonances have been explored in many theories, and there exist different interpretations to these new observed states. Parts of the explorations are reviewed in the following, where many other references have not been mentioned.

In a $^3P_0$ model~\cite{liu3}, $D^*(2600)$ is suggested as an admixture of $2^3S_1$ and $1^3D_1$ with $J^P=1^-$, $D^*(2760)$ is suggested either the
orthogonal partner of $D^*(2600)$ or $1^3D_3$. It is found that $D(2550)$ has a decay width far below the experimental value if it is $2^1S_0$. $D(2750)$ is ignored in the analysis.

In a chiral quark model~\cite{zhong}, $D^*(2600)$ is also suggested as an admixture of $2^3S_1$ and $1^3D_1$ with $J^P=1^-$, $D^*(2760)$ is suggested the $1^3D_3$ with $J^P=3^-$, $D(2750)$ is suggested as an admixture of $1^1D_2$ and $1^3D_2$ with $J^P=2^-$. $D(2550)$ is also found to have a decay width far below the experimental value if it is $2^1S_0$.

In an effective Lagrangian approach based on the heavy quark and chiral symmetry~\cite{wang,colangelo}, $D(2550)$ and $D^*(2600)$ are suggested as the $J^P=(0^-,1^-)$ doublet, $D(2750)$ and $D^*(2760)$ are suggested as the $J^P=(2^-,3^-)$ doublet.

In a heavy-quark effective theory, $D(2550)$ is suggested $2^1S_0$, $D^*(2600)$ is suggested as $2^3S_1$, $D^*(2760)$ is suggested the $1^3D_3$, and $D(2750)$ is suggested as $1D(2^-,{5\over 2})$ or $1D(2^-,{3\over 2})$~\cite{chen2}. $D^*_1(2760)$ is suggested as $1^3D_1$ predominantly, and $D^*_3(2760)$ is regarded as $1^3D_3$~\cite{chen3}.

In the $^3P_0$ model~\cite{gm}, $D(2550)$ and $D^*(2600)$ are regarded as $2^1S_0$ and $2^3S_1$, respectively. $D(2750)$ is regarded as $1D_2$ (mixture of $1^1D_2$ and $1^3D_2$), while $D^*_1(2760)$ and $D^*_3(2760)$ are regarded as $1^1D_1$ and $1^3D_3$, respectively.

In above explorations, $D(2550)$, $D^*(2600)$, $D(2750)$, $D^*_1(2760)$ and $D^*_3(2760)$ have been studied in hadronic decays as initial states. In another way, $D(2550)$, $D^*(2600)$, $D(2750)$, $D^*_1(2760)$ and $D^*_3(2760)$ are possible to be produced in hadronic decays of much more highly excited heavy flavor resonances. In fact, $D(2550)$, $D^*(2600)$, $D(2750)$, $D^*_1(2760)$ and $D^*_3(2760)$ may be identified in their hadronic production. As more and more heavy flavor resonances have been observed in B factories and LHCb facility, these new charmed resonances are expected to be observed as the final states in hadronic decays.

The mass spectrum of radially excited $2D$($2^3D_1$, $2^3D_3$, $2^1D_2$ and $2^3D_2$) $D$ and $D_s$ resonances have been predicted in Refs.~\cite{Zeng,RT,gm}. The hadronic decays of the radially excited $2D$ $D_s$ resonances have been studied in a $^3P_0$ model in Ref.~\cite{Jing Ge}, and more extensive study of highly excited charm and charm-strange mesons has been performed in Ref.~\cite{gm}.

Following our previous work~\cite{Jing Ge}, an extensive study of the hadronic decays of radially excited $2D$ $D$ resonances has been performed in the $^3P_0$ model in this paper. Attention is paid on the hadronic production of $D(2550)$, $D^*(2600)$, $D(2750)$, $D^*_1(2760)$ and $D^*_3(2760)$ from radially excited $2D$ $D$ resonances. In particular, in different assignments, the decay widths and relevant ratios from radially excited $2D$ $D$ resonances to $D(2550)$, $D^*(2600)$, $D(2750)$, $D^*_1(2760)$ and $D^*_3(2760)$ are computed in detail.

\par
The paper is organized as follows. In the second section, a brief review of the $^3P_0$ model is given. Study of hadronic decays of the radially excited $2D$ $D$ mesons is performed in Sec. III. Analysis of $D(2550)$, $D^*(2600)$, $D(2750)$, $D^*_1(2760)$ and $D^*_3(2760)$ is presented in Sec. IV. In Sec. V, conclusions and discussions are given.

\section{A review of the $^3P_0$ model}

\par $^3P_0$ model is known as the quark-pair creation(QPC) model, and has been employed extensively for the calculations of OZI-allowed hadronic decay processes. It was first proposed by Micu~\cite{micu1969} and developed by Yaouanc et al~\cite{yaouanc1,yaouanc2,yaouanc3}. In the model, a pair of $q\bar{q}$ is assumed to be created from the vacuum with $J^{PC}=0^{++}$ and then the created quark/anti-quark regroup with the quarks from the initial meson A to form two daughter mesons B and C. The process of the decay picture is shown in Fig. 1.

\begin{figure}
\begin{center}
\includegraphics[height=2.8cm,angle=0,width=6cm]{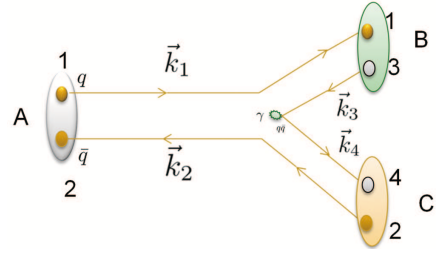}
\caption{Decay process of $A\to B+C$ in the $^3P_0$ model.}
\end{center}
\end{figure}

In $^3P_0$ model, the hadronic decay width $\Gamma$ of $A\to B+C$ is written as,
\begin{eqnarray}
\Gamma  = \pi ^2 \frac{|\vec{k}|}{m_A^2}\sum_{JL} |{\mathcal{M}^{JL}}|^2
\end{eqnarray}
in which $\vec{k}$ is the momentum of the daughter meson in the initial meson A's center of mass frame
\begin{eqnarray}
 |\vec{k}|=\frac{{\sqrt {[m_A^2-(m_B-m_C )^2][m_A^2-(m_B+m_C)^2]}}}{{2m_A }}
\end{eqnarray}
and $\mathcal{M}^{JL}$ is the partial wave amplitude of $A \rightarrow B+C$. In terms of Jacob-Wick formula~\cite{JW}, $\mathcal{M}^{JL}$ can be derived from the helicity amplitude $\mathcal{M}^{M_{J_A } M_{J_B } M_{J_C }}$,
\begin{flalign}
\mathcal{M}^{JL} (A \to BC) &= \frac{{\sqrt {2L + 1} }}{{2J_A  + 1}} \nonumber \\
&\times\sum_{M_{J_B } ,M_{J_C } } \langle {L0JM_{J_A } } |{J_A M_{J_A } }\rangle  \nonumber \\
&\times\langle {J_B M_{J_B } J_C M_{J_C } } |J, {JM_{J_A } } \rangle \nonumber \\
 &\times \mathcal{M}^{M_{J_A } M_{J_B } M_{J_C } } (\vec{K})
\end{flalign}
where $\vec{J}=\vec{J_B}+\vec{J_C}$, $\vec{J_A}=\vec{J_B}+\vec{J_C}+\vec{L}$ and $M_{J_A}=M_{J_B}+M_{J_C}$. In this equation, the helicity amplitude
\begin{flalign}
 &\mathcal{M}^{M_{J_A } M_{J_B } M_{J_C }}\nonumber \\
 &=\sqrt {8E_A E_B E_C } \gamma \sum_{\mbox{\tiny$\begin{array}{c}
M_{L_A } ,M_{S_A } ,\\
M_{L_B } ,M_{S_B } ,\\
M_{L_C } ,M_{S_C } ,m\end{array}$}}  \langle {L_A M_{L_A } S_A M_{S_A } }| {J_A M_{J_A } }\rangle \nonumber \\
 &\times\langle L_B M_{L_B } S_B M_{S_B }|J_B M_{J_B } \rangle \langle L_C M_{L_C } S_C M_{S_C }|J_C M_{J_C }\rangle\nonumber \\
 & \times \langle {1m;1 - m}|{00} \rangle\langle \chi _{S_B M_{S_B }}^{13} \chi _{S_C M_{S_C } }^{24}|\chi _{S_A M_{S_A } }^{12} \chi _{1 - m}^{34}\rangle \nonumber \\
&\times\langle\varphi _B^{13} \varphi _C^{24}|\varphi _A^{12}\varphi _0^{34} \rangle I_{M_{L_B } ,M_{L_C } }^{M_{L_A },m} (\vec{K}).
\end{flalign}
where $I_{M_{L_B },M_{L_C}}^{M_{L_A },m}(\vec{K})$ is the spatial integral
\begin{flalign}
I_{M_{L_B } ,M_{L_C } }^{M_{L_A } ,m} (\vec{K})&= \int d \vec{k}_1 d \vec{k}_2 d \vec{k}_3 d \vec{k}_4 \nonumber \\
&\times\delta ^3 (\vec{k}_1 + \vec{k}_2-\vec{p}_A)\delta ^3 (\vec{k}_3+ \vec{k}_4)\nonumber \\
&\times \delta ^3 (\vec{p}_B- \vec{k}_1- \vec{k}_3 )\delta ^3 (\vec{p}_C- \vec{k}_2 -\vec{k}_4) \nonumber \\
& \times\Psi _{n_B L_B M_{L_B } }^* (\vec{k}_1 ,\vec{k}_3)\Psi _{n_cL_C  M_{L_c}}^* (\vec{k}_2 ,\vec{k}_4) \nonumber \\
& \times \Psi _{n_A L_A M_{LA}} (\vec{k}_1 ,\vec{k}_2 )y _{1m}\left(\frac{\vec{k}_3-\vec{k}_4}{2}\right).
\end{flalign}
In which, $y _{1m}(\vec{k})=|\vec{k}|Y_{1m}(\Omega_{\vec{k}})$ is the solid harmonic polynomial corresponding to the p-wave quark pair, $\vec{k}_{34}=\frac{\vec{k}_3-\vec{k}_4}{2}$ is the momentum of the created quark pair.

In order to compute the flavor matrix element, the following formula is employed,
\begin{flalign}
\langle\varphi _B^{13} \varphi _C^{24}|\varphi _A^{12}\varphi _0^{34} \rangle &= \sum_{I,I^3 } \langle {I_C I_C^3; I_B I^3_B }|{I_A, I_A^3 }\rangle \nonumber \\
&\times [(2I_B  + 1)(2I_C  + 1)(2I_A  + 1)]^{1/2} \nonumber \\
&\times \begin{Bmatrix}
    {I_1} & {I_3} & {I_C}\\
    {I_2} & {I_4} & {I_B}\\
    {I_A} &  {0}  & {I_A}\\ \end{Bmatrix}
\end{flalign}
where $I_i~(i=1,2,3,4)$ is the isospin of the four $u$, $d$, $s$ or $c$ quark. $I_A$, $I_B$ and $I_C$ are the isospins of the mesons $A$, $B$ and $C$, respectively.
$I^3_A$, $I^3_B$ and $I^3_C$ are the isospin third components of the mesons, and $\langle {I_C I_C^3; I_B I^3_B }|{I_A, I_A^3 }\rangle$ is the isospin matrix element.
The spin matrix elements can be written in terms of $9j$ symbols~\cite{yaouanc3},
\begin{flalign}
&\left\langle {\chi _{S_B M_{S_B } }^{13} \chi _{S_C M_{S_C } }^{24} } |{\chi _{S_A M_{S_A } }^{12} \chi _{1 - m}^{34} } \right\rangle \nonumber \\
&= (-1)^{S_C  + 1} [3(2S_B  + 1)(2S_C  + 1)(2S_A  + 1)]^{1/2} \nonumber \\
&\times\sum_{S,M_S } {\langle {S_B M_{S_B } S_C M_{S_C } }|{SM_S }\rangle \langle {SM_S }|{S_A M_{S_A } ;1, - m}\rangle}\nonumber \\
&\times
\begin{Bmatrix}
    {1/2} & {1/2} & {S_B}\\
    {1/2} & {1/2} & {S_C}\\
    {S_A} &  {1}  & {S}\\
\end{Bmatrix}.
\end{flalign}
In $I_{M_{L_B },M_{L_C}}^{M_{L_A },m}(\vec{K})$, a simple harmonic oscillator(SHO) wave function for the mesons is employed as follows,
\begin{flalign}
\Psi_{nLM_L}(\vec{k})&=\frac{(-1)^n(-i)^L}{\beta^{3/2}}\sqrt{\frac{2n!}{\Gamma(n+L+\frac{3}{2})}}\big(\frac{\vec{k}}{\beta}\big)^L exp(-\frac{\vec{k}^2}{2\beta^2}) \nonumber\\
&\times L_n^{L+1/2}\big(\frac{\vec{k}^2}{\beta^2}\big)Y_{LM_L}(\Omega_k)
\end{flalign}
where $L_n^{L+1/2}\big(\frac{\vec{k}^2}{\beta^2}\big)$ denotes the Laguerre polynomial function and $Y_{LM_L}(\Omega_k)$ is a spherical harmonic function. Further details of the indices, matrix elements and other indications in the model can be found in Ref.~\cite{zhang3}. The meson flavor follows the convention in Ref.~\cite{flavor}, $D^0=c\bar{u}$, $D^+=-c\bar{d}$, $D_s^+=-c\bar{s}$, $K^+=-u\bar{s}$, and $K^-=s\bar{u}$.

\section{Hadronic decay of radially excited $2D$ $D$ resonances}

\subsection{Parameters and inputs in $^3P_0$ model}
\par In practical calculation, the numerical results depend on the parameters of the model. Two key parameters in $^3P_0$ model are the quark-pair creation strength $\gamma$ and the harmonic oscillator parameter $\beta$. $\gamma$ indicates the strength of the quark-pair creation from the vacuum and can be obtained by fitting theoretical results of the hadronic decay widths with the experimental data. In Ref.~\cite{TBarnesSGESS} and Ref.~\cite{2750mix2}, the fitting value of $\gamma$ is 0.4 and 0.485. $\gamma=6.947$ is employed in this paper, where $\gamma$ is $\sqrt{96\pi}$ times as that adopted in Ref.~\cite{gm} due to different conventions~\cite{3P0gamma96pi}. For strange quark-pair $s\bar{s}$ creation, $\gamma_{s\bar{s}}=\gamma/\sqrt{3}$~\cite{yaouanc2}.

\par There are often two ways to fix the harmonic oscillator parameter $\beta$. One way is to fix $\beta$ with a common value, which is about $0.4$ GeV~\cite{Ma1,BCPS,HGBSG,054014,6811}. The other way is to fix it with an effective value~\cite{gm,Lx3040,liu2,liu3,054012}. Here, we follow the convention in Ref.~\cite{gm}, where effective values $\beta$ are employed for $S-wave$ , $P-wave$ and $D-wave$ charm and charm strange mesons. These effective SHO $\beta$ values were obtained by equating the root mean square (rms) radius of the SHO wave functions for the specified $(n,l)$ quantum numbers to the rms radius of the wave functions calculated in the relativized quark model~\cite{flavor,gm}, and they are listed in Table I. A common value $\beta=0.4$ GeV is employed for all light flavor mesons as that in Ref.~\cite{gm}.

\par The masses of the constituent quark are taken to be, $m_c=1628$ MeV, $m_u=m_d=220$ MeV, and $m_s=419$ MeV~\cite{gm}. Masses of most mesons used in our calculations are from the Particle Data Group~\cite{pdg}.

For the charm and charm-strange mesons, $m_{D^0}=1864.84$ MeV, $m_{D^\pm}=1869.61$ MeV, $m_{D^{*0}}=2006.97$ MeV, $m_{D^{*\pm}}=2010.27$ MeV, $m_{D(2550)^{0}}=2539.4$ MeV, $m_{D_0^*(2400)^0}=2318.0$ MeV, $m_{D_0^*(2400)^\pm}=2403.0$ MeV, $m_{D_1(2420)^0}=2421.4$ MeV, $m_{D_1(2420)^\pm}=2423.2$ MeV, $m_{D_1(2430)^0}=2427.0$ MeV, $m_{D_2^*(2460)^0}=2462.6$ MeV, $m_{D_2^*(2460)^\pm}=2464.3$ MeV. $m_{D_s^\pm}=1968.3$ MeV, $m_{D_s^{*\pm}}=1968.3$ MeV, $m_{D_{s0}^*(2317)^\pm}=2317.7$ MeV, $m_{D_{s1}(2536)^\pm}=2535.11$ MeV, $m_{D_{s2}(2573)^\pm}=2571.9$ MeV, $m_{D_{s1}^*(2700)^\pm}=2709.0$ MeV.

For light flavor mesons, $m_{\pi^0}=134.977$ MeV, $m_{\pi^\pm}=139.570$ MeV, $m_{K^0}=497.611$ MeV, $m_{K^\pm}=493.677$ MeV, $m_{\rho(770)^0}=775.26$ MeV, $m_{\rho(770)^\pm}=775.11$ MeV, $m_{\pi(1300)^{0,\pm}}=1300.0$ MeV, $m_{b_1(1235)^{0,\pm}}=1229.5$ MeV, $m_{a_1(1260)^{0,\pm}}=1255.0$ MeV, $m_{a_2(1320)^{0,\pm}}=1318.3$ MeV. $m_{\eta}=547.862$ MeV, $m_{\eta^\prime}=957.78$ MeV, $m_{\phi(1020)}=1019.461$ MeV, $m_{\omega}=782.65$ MeV, $m_{K^*(892)^0}=895.81$ MeV, $m_{K^*(892)^\pm}=891.66$ MeV.

For resonances not observed by experiment, their masses are taken from theoretical predictions in Ref.~\cite{gm}, $m_{D(2^3D_1)}=3231$ MeV, $m_{D(2^3D_3)}=3226$ MeV, $m_{D(2D_2)}=3212$ MeV, $m_{D(2D_2^\prime)}=3248$ MeV, $m_{D(2P_1)}=2924$ MeV, $m_{D(2P_1^\prime)}=2961$ MeV, $m_{D(2^3P_0)}=2931$ MeV, $m_{D(2^3P_2)}=2957$ MeV, $m_{D_s(2^1S_0)}=2673$ MeV.

Masses of $D^*(2600)$, $D(2750)$, $D_1^*(2760)$ and $D_3^*(2760)$ come from the experimental data in Ref~\cite{Benitez,lhcb1,lhcb2,lhcb3}, $m_{D^*(2600)^0}=2628.95$ MeV (mean value), $m_{D^*(2600)^+}=2621.3$ MeV, $m_{D(2750)^0}=2744.7$ MeV (mean value), $m_{D_1^*(2760)^{0,\pm}}=2781$ MeV, $m_{D_3^*(2760)^{0,\pm}}=2798$ MeV.

\begin{table}[t]
\caption{States, masses and $\beta$ values of the mesons (unit in MeV).}
\begin{tabular}{p{0.0cm}p{0.9cm}p{1.8cm}p{1.1cm}p{0.9cm}p{1.8cm}p{1.1cm}}
   \hline\hline
   &$States     $ & $Name                $ & $\beta$ & $States     $ & $Name                $ & $\beta$\\
   \hline
   &$1^1S_0     $ & $D^{0,\pm}           $ & $601  $ & $1^1S_0     $ & $D_s^{0,\pm}         $ & $651  $\\
   &$1^3S_1     $ & $D^{*0,\pm}          $ & $516  $ & $1^3S_1     $ & $D_s^{*,\pm}         $ & $562  $\\
   &$2^1S_0     $ & $D(2550)^{0,\pm}     $ & $450  $ & $2^1S_0     $ & $D_s(2^1S_0)         $ & $475  $\\
   &$2^3S_1     $ & $D(2^3S_1)           $ & $434  $ & $2^3S_1     $ & $D_{s1}^*(2700)^{\pm}$ & $458  $\\

   &$1^3P_0     $ & $D_0^*(2400)^{0,\pm} $ & $516  $ & $1^3P_0     $ & $D_{s0}^*(2317)^{\pm}$ & $542  $\\
   &$1P_1       $ & $D_1(2420)^{0,\pm}   $ & $475  $ & $           $ & $                    $ & $     $\\
   &$1P_1^\prime$ & $D_1(2430)^{0,\pm}   $ & $475  $ & $           $ & $                    $ & $     $\\
   &$1^3P_2     $ & $D_2^*(2460)^{0,\pm} $ & $437  $ & $1^3P_2     $ & $D_{s2}(2573)^{\pm}  $ & $464  $\\

   &$2^3P_0     $ & $D(2^3P_0)           $ & $431  $ & $           $ & $                    $ & $     $\\
   &$2P_1       $ & $D(2P_1)             $ & $417  $ & $           $ & $                    $ & $     $\\
   &$2P_1^\prime$ & $D(2P_1^\prime)      $ & $417  $ & $           $ & $                    $ & $     $\\
   &$2^3P_2     $ & $D(2^3P_2)           $ & $402  $ & $           $ & $                    $ & $     $\\

   &$1^3D_1     $ & $D(1^3D_1)           $ & $456  $ & $           $ & $                    $ & $     $\\
   &$1D_2       $ & $D(1D_2)             $ & $428  $ & $           $ & $                    $ & $     $\\
   &$1D_2^\prime$ & $D(1D_2)^\prime      $ & $428  $ & $           $ & $                    $ & $     $\\
   &$1^3D_3     $ & $D(1^3D_3)           $ & $407  $ & $           $ & $                    $ & $     $\\

   &$2^3D_1     $ & $D(2^3D_1)           $ & $410  $ & $           $ & $                    $ & $     $\\
   &$2D_2       $ & $D(2D_2)             $ & $396  $ & $           $ & $                    $ & $     $\\
   &$2D_2^\prime$ & $D(2D_2)^\prime      $ & $396  $ & $           $ & $                    $ & $     $\\
   &$2^3D_3     $ & $D(2^3D_3)           $ & $385  $ & $           $ & $                    $ & $     $\\

   \hline\hline
\end{tabular}
\label{table2}
\end{table}

\subsection{OZI-allowed decay modes and channels.}
For radially excited $2D$ $D$ resonances, there exist four resonances: $2^3D_1$, $2^3D_3$, $2D_2$ and $2D^\prime_2$~\cite{gm}. Due to conservation of $J^P$ quantum numbers and kinematical permission, there are $12$ OZI-allowed decay modes for each resonance, and each mode contains a few decay channels. All possible kinematically permitted and OZI-allowed decay modes and decay channels of these four radially excited $2D~D$ resonances are presented in from Table II to Table V. In each table, all the decay modes and channels of the $2D~D$ resonances with $c\bar{u}$ and $c\bar{d}$ flavors are given. Some decay channels ignored in Ref.~\cite{gm} are also presented in these tables.

\begin{table*}[]
\caption{OZI-allowed hadronic decay modes and channels of $D(2^3D_1)$.}
\begin{tabular}{p{0cm}p{1.2cm}p{6.5cm}p{1.2cm}p{6.5cm}}
   \hline\hline
   &&$D(2^3D_1)$ as $c\bar{u}$ &&$D(2^3D_1)$ as $c\bar{d}$\\
   \hline
   &Mode &Channels &Mode &Channels\\
   \hline
    & $0^{-}+0^{-}$  & $D^0\pi^0,D^+\pi^-,D_s^+K^-, D^0\eta, D^0\eta^\prime,$                & $0^{-}+0^{-}$  & $D^0\pi^+, D^+\pi^0, D_s^+K^0, D^+\eta, D^+\eta^\prime,$ \\
    & $           $  & $D^0\pi(1300)^0, D^+\pi(1300)^-,$                                     & $           $  & $D^0\pi(1300)^+, D^+\pi(1300)^0,$ \\
    & $           $  & $D(2550)^0\pi^0, D(2550)^+\pi^-, D(2550)^0\eta,       $               & $           $  & $D(2550)^0\pi^+, D(2550)^+\pi^0, D(2550)^+\eta,$\\
    & $           $  & $D_s(2^1S_0)^+K^-                           $                         & $           $  & $ D_s(2^1S_0)^+K^0 $ \\
                  & $ $  & $  $      & $           $  & $        $ \\

    & $0^{-}+1^{-}$  & $D^0\rho(770)^0, D^+\rho(770)^-, D^0\phi(1020), D^0\omega$            & $0^{-}+1^{-}$  & $D^{0}\rho(770)^+, D^+\rho(770)^0, D^+\phi(1020), D^+\omega $ \\
    & $           $  & $D_s^+K^*(892)^-$                                                     & $           $  & $D_s^+K^*(892)^0 $ \\
                  & $ $  & $  $      & $           $  & $        $ \\

    & $0^{-}+1^{+}$  & $D^0b_1(1235)^0, D^+b_1(1235)^-, $                                    & $0^{-}+1^{+}$  & $D^0b_1(1235)^+,D^+b_1(1235)^0, $\\
    & $           $  & $D^0a_1(1260)^0, D^+a_1(1260)^-                         $             & $           $  & $D^0a_1(1260)^+, D^+a_1(1260)^0  $   \\
                  & $ $  & $  $      & $           $  & $        $ \\

    & $0^{-}+2^{+}$  & $D^0a_2(1320)^0, D^+a_2(1320)^-$                                      & $0^{-}+2^{+}$  & $D^0a_2(1320)^+, D^+a_2(1320)^0$\\
                  & $ $  & $  $      & $           $  & $        $ \\

    & $1^{-}+0^{-}$ & $D^{*0}\pi^0, D^{*+}\pi^-, D^{*0}\eta, D^{*0}\eta^\prime, D_s^{*+}K^-,$& $1^{-}+0^{-}$  & $D^{*0}\pi^+, D^{*+}\pi^0, D^{*+}\eta, D^{*+}\eta^\prime, D_s^{*+}K^0,$\\
    & $           $  & $D(1^3D_1)^0\pi^0, D(1^3D_1)^+\pi^-, D_{s1}^*(2700)^+K^-,$            & $           $  & $D(1^3D_1)^0\pi^+, D(1^3D_1)^+\pi^0, D_{s1}^*(2700)^+K^0,$\\
    & $           $  & $D(2^3S_1)^0\pi^0, D(2^3S_1)^+\pi^-$                                  & $           $  & $D(2^3S_1)^0\pi^+, D(2^3S_1)^+\pi^0$ \\
                  & $ $  & $  $      & $           $  & $        $ \\

    & $1^{-}+1^{-}$  & $D^{*0}\rho(770)^0, D^{*+}\rho(770)^-, D^{*0}\phi(1020), $            & $1^{-}+1^{-}$  & $D^{*0}\rho(770)^+, D^{*+}\rho(770)^0, D^{*+}\phi(1020),$ \\
    & $           $  & $D^{*0}\omega, D_s^{*+}K^*(892)^-$                                    & $           $  & $D^{*+}\omega, D_s^{*+}K^*(892)^0$ \\
                  & $ $  & $  $      & $           $  & $        $ \\

    & $0^{+}+1^{-}$  & $D^*_0(2400)^0\rho(770)^0, D^*_0(2400)^+\rho(770)^-,  $               & $0^{+}+1^{-}$  & $D^*_0(2400)^0\rho(770)^+, D^*_0(2400)^+\rho(770)^0, $\\
    & $           $  & $D^*_0(2400)^0\omega, D^*_{s0}(2317)^{+}K^*(892)^- $                  & $           $  & $D^*_0(2400)^+\omega, D^*_{s0}(2317)^{+}K^*(892)^0  $   \\
                  & $ $  & $  $      & $           $  & $        $ \\

    & $1^{+}+0^{-}$  & $D_1(2420)^0\pi^{0}, D_1(2420)^+\pi^{-}, D_1(2420)^0\eta,$            & $1^{+}+0^{-}$  & $D_1(2420)^0\pi^{+}, D_1(2420)^+\pi^{0}, D_1(2420)^+\eta,$\\
    & $           $  & $D_{s1}(2536)^+K^-,$                                                  & $           $  & $D_{s1}(2536)^+K^0,$\\
    & $           $  & $D_1(2430)^0\pi^{0}, D_1(2430)^+\pi^{-}, D_1(2430)^0\eta,$            & $           $  & $D_1(2430)^0\pi^{+}, D_1(2430)^+\pi^{0}, D_1(2430)^+\eta,$\\
    & $           $  & $D_{s1}(2460)^{+}K^-$                                                 & $           $  & $D_{s1}(2460)^{+}K^0,$\\
    & $           $  & $D(2P_1)^0\pi^{0}, D(2P_1)^+\pi^{-} $                                 & $           $  & $D(2P_1)^0\pi^{+}, D(2P_1)^+\pi^{0}$\\
    & $           $  & $D(2P_1^\prime)^0\pi^{0}, D(2P_1^\prime)^+\pi^{-} $                   & $           $  & $D(2P_1^\prime)^0\pi^{+}, D(2P_1^\prime)^+\pi^{0}$\\
                  & $ $  & $  $      & $           $  & $        $ \\

    & $1^{+}+1^{-}$  & $D_1(2420)^0\rho(770)^0, D_1(2420)^+\rho(770)^-,$                     & $1^{+}+1^{-}$  & $D_1(2420)^0\rho(770)^+, D_1(2420)^+\rho(770)^0,$ \\
    & $           $  & $D_1(2420)^0\omega,$                                                  & $           $  & $D_1(2420)^+\omega,$ \\
    & $           $  & $D_1(2430)^0\rho(770)^0, D_1(2430)^+\rho(770)^-, $                    & $           $  & $D_1(2430)^0\rho(770)^+, D_1(2430)^+\rho(770)^0,$ \\
    & $           $  & $D_1(2430)^0\omega $                                                  & $           $  & $D_1(2430)^+\omega,$ \\
                  & $ $  & $  $      & $           $  & $        $ \\

    & $2^{-}+0^{-}$  & $(1^1D_2)\pi^0, (1^1D_2)\pi^-, $                                      & $2^{-}+0^{-}$  & $(1^1D_2)\pi^+, (1^1D_2)\pi^0,$\\
    & $           $  & $(1^3D_2)\pi^0, (1^3D_2)\pi^- $                                       &                & $(1^3D_2)\pi^+, (1^3D_2)\pi^0 $\\
                  & $ $  & $  $      & $           $  & $        $ \\

    & $2^{+}+0^{-}$  & $D_2^*(2460)^0\pi^0, D_2^*(2460)^+\pi^-, D_2^*(2460)^0\eta,$          & $2^{+}+0^{-}$  & $D_2^*(2460)^0\pi^+, D_2^*(2460)^+\pi^0, D_2^*(2460)^+\eta,$\\
    & $           $  & $D_{s2}^*(2573)^+K^-, D(2^3P_2)^0\pi^0, D(2^3P_2)^+\pi^-$             &                & $D_{s2}^*(2573)^+K^0, D(2^3P_2)^0\pi^+, D(2^3P_2)^+\pi^0 $\\
                  & $ $  & $  $      & $           $  & $        $ \\

    & $3^{-}+0^{-}$  &$D(1^3D_3)^0\pi^0,D(1^3D_3)^+\pi^-$                                    & $3^{-}+0^{-}$  & $D(1^3D_3)^0\pi^+,D(1^3D_3)^+\pi^0 $\\
                  & $ $  & $  $      & $           $  & $        $ \\

 \hline\hline
\end{tabular}
\label{table3}
\end{table*}

\begin{table*}[]
\caption{OZI-allowed hadronic decay modes and channels of $D(2^3D_3)$.}
\begin{tabular}{p{0cm}p{1.2cm}p{6.5cm}p{1.2cm}p{6.5cm}}
   \hline\hline
   &&$D(2^3D_3)$ as $c\bar{u}$ &&$D(2^3D_3)$ as $c\bar{d}$\\
   \hline
   &Mode &Channels &Mode &Channels\\
   \hline
    & $0^{-}+0^{-}$  & $D^0\pi^0,D^+\pi^-,D_s^+K^-, D^0\eta, D^0\eta^\prime,$                & $0^{-}+0^{-}$  & $D^0\pi^+, D^+\pi^0, D_s^+K^0, D^+\eta, D^+\eta^\prime,$ \\
    & $           $  & $D^0\pi(1300)^0, D^+\pi(1300)^-,$                                     & $           $  & $D^0\pi(1300)^+, D^+\pi(1300)^0,$ \\
    & $           $  & $D(2550)^0\pi^0, D(2550)^+\pi^-, D(2550)^0\eta,       $               & $           $  & $D(2550)^0\pi^+, D(2550)^+\pi^0, D(2550)^+\eta,$\\
    & $           $  & $D_s(2^1S_0)^+K^-                           $                         & $           $  & $ D_s(2^1S_0)^+K^0 $ \\
                  & $ $  & $  $      & $           $  & $        $ \\

    & $0^{-}+1^{-}$  & $D^0\rho(770)^0, D^+\rho(770)^-, D^0\phi(1020), D^0\omega$            & $0^{-}+1^{-}$  & $D^{0}\rho(770)^+, D^+\rho(770)^0, D^+\phi(1020), D^+\omega $ \\
    & $           $  & $D_s^+K^*(892)^-$ \\
                  & $ $  & $  $      & $           $  & $        $ \\

    & $0^{-}+1^{+}$  & $D^0b_1(1235)^0, D^+b_1(1235)^-, $                                    & $0^{-}+1^{+}$  & $D^0b_1(1235)^+,D^+b_1(1235)^0, $\\
    & $           $  & $D^0a_1(1260)^0, D^+a_1(1260)^-                         $             & $           $  & $D^0a_1(1260)^+, D^+a_1(1260)^0  $   \\
                  & $ $  & $  $      & $           $  & $        $ \\

    & $0^{-}+2^{+}$  &$D^0a_2(1320)^0,D^+a_2(1320)^-$                                        & $0^{-}+2^{+}$  & $D^0a_2(1320)^+,D^+a_2(1320)^0$\\
                  & $ $  & $  $      & $           $  & $        $ \\

    & $1^{-}+0^{-}$ & $D^{*0}\pi^0, D^{*+}\pi^-, D^{*0}\eta, D^{*0}\eta^\prime, D_s^{*+}K^-,$& $1^{-}+0^{-}$  & $D^{*0}\pi^+, D^{*+}\pi^0, D^{*+}\eta, D^{*+}\eta^\prime, D_s^{*+}K^0,$\\
    & $           $  & $D(1^3D_1)^0\pi^0, D(1^3D_1)^+\pi^-, D_{s1}^*(2700)^+K^-,$            & $           $  & $D(1^3D_1)^0\pi^+, D(1^3D_1)^+\pi^0, D_{s1}^*(2700)^+K^0,$\\
    & $           $  & $D(2^3S_1)^0\pi^0, D(2^3S_1)^+\pi^- $                                 & $           $  & $D(2^3S_1)^0\pi^+, D(2^3S_1)^+\pi^0$ \\
                  & $ $  & $  $      & $           $  & $        $ \\

    & $1^{-}+1^{-}$  & $D^{*0}\rho(770)^0, D^{*+}\rho(770)^-, D^{*0}\phi(1020), $            & $1^{-}+1^{-}$  & $D^{*0}\rho(770)^+, D^{*+}\rho(770)^0, D^{*+}\phi(1020),$ \\
    & $           $  & $D^{*0}\omega, D_s^{*+}K^*(892)^-$                                    & $           $  & $D^{*+}\omega, D_s^{*+}K^*(892)^0$ \\
                  & $ $  & $  $      & $           $  & $        $ \\



    & $0^{+}+1^{-}$  & $D^*_0(2400)^0\rho(770)^0, D^*_0(2400)^+\rho(770)^-,  $               & $0^{+}+1^{-}$  & $D^*_0(2400)^0\rho(770)^+, D^*_0(2400)^+\rho(770)^0, $\\
    & $           $  & $D^*_0(2400)^0\omega, D^*_{s0}(2317)^{+}K^*(892)^- $                  & $           $  & $D^*_0(2400)^+\omega, D^*_{s0}(2317)^{+}K^*(892)^0  $   \\
                  & $ $  & $  $      & $           $  & $        $ \\

    & $1^{+}+0^{-}$  & $D_1(2420)^0\pi^{0}, D_1(2420)^+\pi^{-}, D_1(2420)^0\eta,$            & $1^{+}+0^{-}$  & $D_1(2420)^0\pi^{+}, D_1(2420)^+\pi^{0}, D_1(2420)^+\eta,$\\
    & $           $  & $D_{s1}(2536)^+K^-,$                                                  & $           $  & $D_{s1}(2536)^+K^0,$\\
    & $           $  & $D_1(2430)^0\pi^{0}, D_1(2430)^+\pi^{-}, D_1(2430)^0\eta,$            & $           $  & $D_1(2430)^0\pi^{+}, D_1(2430)^+\pi^{0}, D_1(2430)^+\eta,$\\
    & $           $  & $D_{s1}(2460)^{+}K^-$                                                 & $           $  & $D_{s1}(2460)^{+}K^0,$\\
    & $           $  & $D(2P_1)^0\pi^{0}, D(2P_1)^+\pi^{-} $                                 & $           $  & $D(2P_1)^0\pi^{+}, D(2P_1)^+\pi^{0}$\\
    & $           $  & $D(2P_1^\prime)^0\pi^{0}, D(2P_1^\prime)^+\pi^{-} $                   & $           $  & $D(2P_1^\prime)^0\pi^{+}, D(2P_1^\prime)^+\pi^{0}$\\
                  & $ $  & $  $      & $           $  & $        $ \\

    & $1^{+}+1^{-}$  & $D_1(2420)^0\rho(770)^0, D_1(2420)^+\rho(770)^-,$                     & $1^{+}+1^{-}$  & $D_1(2420)^0\rho(770)^+, D_1(2420)^+\rho(770)^0,$ \\
    & $           $  & $D_1(2420)^0\omega,$                                                  & $           $  & $D_1(2420)^+\omega,$ \\
    & $           $  & $D_1(2430)^0\rho(770)^0, D_1(2430)^+\rho(770)^-, $                    & $           $  & $D_1(2430)^0\rho(770)^+, D_1(2430)^+\rho(770)^0,$ \\
    & $           $  & $D_1(2430)^0\omega $                                                  & $           $  & $D_1(2430)^+\omega,$ \\
                  & $ $  & $  $      & $           $  & $        $ \\

    & $2^{-}+0^{-}$  &$(1^1D_2)D(2750)^0\pi^0,(1^1D_2)D(2750)^+\pi^-, $                      & $2^{-}+0^{-}$  & $(1^1D_2)D(2750)^0\pi^+,(1^1D_2)D(2750)^+\pi^0,$\\
    & $           $  &$(1^3D_2)D(2750)^0\pi^0,(1^3D_2)D(2750)^+\pi^- $                       &                & $(1^3D_2)D(2750)^0\pi^+,(1^3D_2)D(2750)^+\pi^0 $\\
                  & $ $  & $  $      & $           $  & $        $ \\

    & $2^{+}+0^{-}$  &$D_2^*(2460)^0\pi^0,D_2^*(2460)^+\pi^-,D_{s2}^*(2573)^+K^-,$           & $2^{+}+0^{-}$  & $D_2^*(2460)^0\pi^+,D_2^*(2460)^+\pi^0,D_{s2}^*(2573)^+K^0,$\\
    & $           $  &$D(2^3P_2)^0\pi^0,D(2^3P_2)^+\pi^-$                                    &                & $D(2^3P_2)^0\pi^+,D(2^3P_2)^+\pi^0 $\\
                  & $ $  & $  $      & $           $  & $        $ \\


    & $3^{-}+0^{-}$  &$D(1^3D_3)^0\pi^0,D(1^3D_3)^+\pi^-$                                    & $3^{-}+0^{-}$  & $D(1^3D_3)^0\pi^+,D(1^3D_3)^+\pi^0 $\\
                  & $ $  & $  $      & $           $  & $        $ \\
     \hline\hline
\end{tabular}
\label{table3}
\end{table*}

\begin{table*}[]
\caption{OZI-allowed hadronic decay modes and channels of $D(2 D_2)$.}
\begin{tabular}{p{0cm}p{1.2cm}p{6.5cm}p{1.2cm}p{6.5cm}}
   \hline\hline
   &&$D(2 D_2)$ as $c\bar{u}$ &&$D(2 D_2)$ as $c\bar{d}$\\
   \hline
   &Mode &Channels &Mode &Channels\\
   \hline
    & $0^{-}+1^{-}$  & $D^0\rho(770)^0, D^+\rho(770)^-, D^0\phi(1020), D^0\omega,$           & $0^{-}+1^{-}$  & $D^{0}\rho(770)^+, D^+\rho(770)^0, D^+\phi(1020), D^+\omega,$ \\
    & $           $  & $D_s^+K^*(892)^-$                                                     & $           $  & $D_s^+K^*(892)^0 $ \\
                  & $ $  & $  $      & $           $  & $        $ \\

    & $0^{-}+1^{+}$  & $D^0b_1(1235)^0, D^+b_1(1235)^-,                    $                 & $0^{-}+1^{+}$  & $D^0b_1(1235)^+, D^+b_1(1235)^0,$\\
    & $           $  & $D^0a_1(1260)^0, D^+a_1(1260)^-                     $                 & $           $  & $D^0a_1(1260)^+, D^+a_1(1260)^0  $   \\
                  & $ $  & $  $      & $           $  & $        $ \\

    & $0^{-}+2^{+}$  &$D^0a_2(1320)^0, D^+a_2(1320)^-$                                       & $0^{-}+2^{+}$  & $D^0a_2(1230)^+, D^+a_2(1320)^0$\\
                  & $ $  & $  $      & $           $  & $        $ \\

    & $1^{-}+0^{-}$  & $D^{*0}\pi^0, D^{*+}\pi^-, D^{*0}\eta, D^{*0}\eta^\prime, $           & $1^{-}+0^{-}$  & $D^{*0}\pi^+, D^{*+}\pi^0, D^{*+}\eta, D^{*+}\eta^\prime,$\\
    & $           $  & $D_s^{*+}K^-, $                                                       & $           $  & $D_s^{*+}K^0,$\\
    & $           $  & $D(1^3D_1)^0\pi^0, D(1^3D_1)^+\pi^-, D_{s1}^*(2700)^+K^-,$            & $           $  & $D(1^3D_1)^0\pi^+, D(1^3D_1)^+\pi^0, D_{s1}^*(2700)^+K^0,$\\
    & $           $  & $D(2^3S_1)^0\pi^0, D(2^3S_1)^+\pi^-$                                  & $           $  & $D(2^3S_1)^0\pi^+, D(2^3S_1)^+\pi^0$ \\
                  & $ $  & $  $      & $           $  & $        $ \\

    & $1^{-}+1^{-}$  &$D^{*0}\rho(770)^0, D^{*+}\rho(770)^-, D^{*0}\phi(1020),$              & $1^{-}+1^{-}$  & $D^{*0}\rho(770)^+, D^{*+}\rho(770)^0, D^{*+}\phi(1020),$ \\
    & $           $  &$D^{*0}\omega, D_s^{*+}K^*(892)^-$                                     & $           $  & $D^{*+}\omega, D_s^{*+}K^*(892)^0$ \\
                  & $ $  & $  $      & $           $  & $        $ \\


    & $0^{+}+0^{-}$  & $D_0^*(2400)^0\pi^0, D_0^*(2400)^+\pi^-, D_0^*(2400)^0\eta,$          & $0^{+}+0^{-}$  & $D_0^*(2400)^0\pi^+, D_0^*(2400)+\pi^0, D_0^*(2400)^+\eta,$ \\
    & $           $  & $D_{s0}^*(2317)^{+}K^-,$                                              & $           $  & $D_{s0}^*(2317)^{+}K^0, $ \\
    & $           $  & $D(2^3P_0)^0\pi^0, D(2^3P_0)^+\pi^-$                                  & $           $  & $D(2^3P_0)^0\pi^+, D(2^3P_0)^+\pi^0$ \\
                  & $ $  & $  $      & $           $  & $        $ \\

    & $0^{+}+1^{-}$  & $D^*_0(2400)^0\rho(770)^0, D^*_0(2400)^+\rho(770)^-,$                 & $0^{+}+1^{-}$  & $D^*_0(2400)^0\rho(770)^+, D^*_0(2400)^+\rho(770)^0,$\\
    & $           $  & $D^*_0(2400)^0\omega, D^*_s0(2317)^{+}K^*(892)^-$                     & $           $  & $D^*_0(2400)^+\omega, D^*_s0(2317)^{+}K^*(892)^0$ \\
                  & $ $  & $  $      & $           $  & $        $ \\

    & $1^{+}+0^{-}$  & $D_1(2420)^0\pi^{0}, D_1(2420)^+\pi^{-}, D_1(2420)^0\eta,$            & $1^{+}+0^{-}$  & $D_1(2420)^0\pi^{+}, D_1(2420)^+\pi^{0}, D_1(2420)^+\eta,$\\
    & $           $  & $D_{s1}(2536)^+K^-$                                                   & $           $  & $D_{s1}(2536)^+K^0,$\\
    & $           $  & $D_1(2430)^0\pi^{0}, D_1(2430)^+\pi^{-}, D_1(2430)^0\eta,$            & $           $  & $D_1(2430)^0\pi^{+}, D_1(2430)^+\pi^{0}, D_1(2430)^+\eta,$\\
    & $           $  & $D_{s1}(2460)^{+}K^-$                                                 & $           $  & $D_{s1}(2460)^{+}K^0,$\\
    & $           $  & $D(2P_1)^0\pi^{0}, D(2P_1)^+\pi^{-} $                                 & $           $  & $D(2P_1)^0\pi^{+}, D(2P_1)^+\pi^{0}$\\
    & $           $  & $D(2P_1^\prime)^0\pi^{0}, D(2P_1^\prime)^+\pi^{-} $                   & $           $  & $D(2P_1^\prime)^0\pi^{+}, D(2P_1^\prime)^+\pi^{0}$\\
                  & $ $  & $  $      & $           $  & $        $ \\

    & $1^{+}+1^{-}$  & $D_1(2420)^0\rho(770)^0, D_1(2420)^+\rho(770)^-,$                     & $1^{+}+1^{-}$  & $D_1(2420)^0\rho(770)^+, D_1(2420)^+\rho(770)^0,$ \\
    & $           $  & $D_1(2420)^0\omega,$                                                  & $           $  & $D_1(2420)^+\omega,$ \\
    & $           $  & $D_1(2430)^0\rho(770)^0, D_1(2430)^+\rho(770)^-, $                    & $           $  & $D_1(2430)^0\rho(770)^+, D_1(2430)^+\rho(770)^0,$ \\
    & $           $  & $D_1(2430)^0\omega $                                                  & $           $  & $D_1(2430)^+\omega $ \\
                  & $ $  & $  $      & $           $  & $        $ \\

    & $2^{+}+0^{-}$  &$D_2^*(2460)^0\pi^0, D_2^*(2460)^+\pi^-, D_{s2}^*(2573)^+K^-,$         & $2^{+}+0^{-}$  & $D_2^*(2460)^0\pi^+, D_2^*(2460)^+\pi^0, D_{s2}^*(2573)^+K^0,$\\
    & $           $  &$D(2^3P_2)^0\pi^0, D(2^3P_2)^+\pi^-$                                   &                & $D(2^3P_2)^0\pi^+, D(2^3P_2)^+\pi^0 $\\
                  & $ $  & $  $      & $           $  & $        $ \\


    & $2^{-}+0^{-}$  &$D(1^1D_2)\pi^0, (1^1D_2)D(2750)^+\pi^-, $                             & $2^{-}+0^{-}$  & $D(1^1D_2)\pi^+, (1^1D_2)D(2750)^+\pi^0,$\\
    & $           $  &$D(1^3D_2)\pi^0, (1^3D_2)D(2750)^+\pi^- $                              &                & $D(1^3D_2)\pi^+, (1^3D_2)D(2750)^+\pi^0 $\\
                  & $ $  & $  $      & $           $  & $        $ \\

    & $3^{-}+0^{-}$  &$D(1^3D_3)^0\pi^0, D(1^3D_3)^+\pi^-$                                   & $3^{-}+0^{-}$  & $D(1^3D_3)^0\pi^+, D(1^3D_3)^+\pi^0 $\\
                  & $ $  & $  $      & $           $  & $        $ \\
     \hline\hline
\end{tabular}
\label{table3}
\end{table*}

\begin{table*}[]
\caption{OZI-allowed hadronic decay modes and channels of $D(2 D_2^\prime)$.}
\begin{tabular}{p{0cm}p{1.2cm}p{6.5cm}p{1.2cm}p{6.5cm}}
   \hline\hline
   &&$D(2 D_2)$ as $c\bar{u}$ &&$D(2 D_2)$ as $c\bar{d}$\\
   \hline
   &Mode &Channels &Mode &Channels\\
   \hline
    & $0^{-}+1^{-}$  & $D^0\rho(770)^0, D^+\rho(770)^-, D^0\phi(1020), D^0\omega,$           & $0^{-}+1^{-}$  & $D^{0}\rho(770)^+, D^+\rho(770)^0, D^+\phi(1020), D^+\omega,$ \\
    & $           $  & $D_s^+K^*(892)^-$                                                     & $           $  & $D_s^+K^*(892)^0 $ \\
                  & $ $  & $  $      & $           $  & $        $ \\
    & $0^{-}+1^{+}$  & $D^0b_1(1235)^0, D^+b_1(1235)^-,                    $                 & $0^{-}+1^{+}$  & $D^0b_1(1235)^+, D^+b_1(1235)^0,$\\
    & $           $  & $D^0a_1(1260)^0, D^+a_1(1260)^-                     $                 & $           $  & $D^0a_1(1260)^+, D^+a_1(1260)^0  $   \\
                  & $ $  & $  $      & $           $  & $        $ \\

    & $0^{-}+2^{+}$  &$D^0a_2(1320)^0, D^+a_2(1320)^-$                                       & $0^{-}+2^{+}$  & $D^0a_2(1230)^+, D^+a_2(1320)^0$\\
                  & $ $  & $  $      & $           $  & $        $ \\

    & $1^{-}+0^{-}$  & $D^{*0}\pi^0, D^{*+}\pi^-, D^{*0}\eta, D^{*0}\eta^\prime, $           & $1^{-}+0^{-}$  & $D^{*0}\pi^+, D^{*+}\pi^0, D^{*+}\eta, D^{*+}\eta^\prime,$\\
    & $           $  & $D_s^{*+}K^-, $                                                       & $           $  & $D_s^{*+}K^0,$\\
    & $           $  & $D(1^3D_1)^0\pi^0, D(1^3D_1)^+\pi^-, D_{s1}^*(2700)^+K^-,$            & $           $  & $D(1^3D_1)^0\pi^+, D(1^3D_1)^+\pi^0, D_{s1}^*(2700)^+K^0,$\\
    & $           $  & $D(2^3S_1)^0\pi^0, D(2^3S_1)^+\pi^-, D_s(2^3S_1)^+K^- $               & $           $  & $D(2^3S_1)^0\pi^+, D(2^3S_1)^+\pi^0, D_s(2^3S_1)^+K^0$ \\
                  & $ $  & $  $      & $           $  & $        $ \\

    & $1^{-}+1^{-}$  &$D^{*0}\rho(770)^0, D^{*+}\rho(770)^-, D^{*0}\phi(1020),$              & $1^{-}+1^{-}$  & $D^{*0}\rho(770)^+, D^{*+}\rho(770)^0, D^{*+}\phi(1020),$ \\
    & $           $  &$D^{*0}\omega, D_s^{*+}K^*(892)^-$                                     & $           $  & $D^{*+}\omega, D_s^{*+}K^*(892)^0$ \\
                  & $ $  & $  $      & $           $  & $        $ \\\


    & $0^{+}+0^{-}$  & $D_0^*(2400)^0\pi^0, D_0^*(2400)^+\pi^-, D_0^*(2400)^0\eta,$          & $0^{+}+0^{-}$  & $D_0^*(2400)^0\pi^+, D_0^*(2400)+\pi^0, D_0^*(2400)^+\eta,$ \\
    & $           $  & $D_{s0}^*(2317)^{+}K^-,$                                              & $           $  & $D_{s0}^*(2317)^{+}K^0, $ \\
    & $           $  & $D(2^3P_0)^0\pi^0, D(2^3P_0)^+\pi^-$                                  & $           $  & $D(2^3P_0)^0\pi^+, D(2^3P_0)^+\pi^0$ \\
                  & $ $  & $  $      & $           $  & $        $ \\

    & $0^{+}+1^{-}$  & $D^*_0(2400)^0\rho(770)^0, D^*_0(2400)^+\rho(770)^-,$                 & $0^{+}+1^{-}$  & $D^*_0(2400)^0\rho(770)^+, D^*_0(2400)^+\rho(770)^0,$\\
    & $           $  & $D^*_0(2400)^0\omega, D^*_s0(2317)^{+}K^*(892)^-$                     & $           $  & $D^*_0(2400)^+\omega, D^*_s0(2317)^{+}K^*(892)^0$ \\
                  & $ $  & $  $      & $           $  & $        $ \\

    & $1^{+}+0^{-}$  & $D_1(2420)^0\pi^{0}, D_1(2420)^+\pi^{-}, D_1(2420)^0\eta,$            & $1^{+}+0^{-}$  & $D_1(2420)^0\pi^{+}, D_1(2420)^+\pi^{0}, D_1(2420)^+\eta,$\\
    & $           $  & $D_{s1}(2536)^+K^-,$                                                   & $           $  & $D_{s1}(2536)^+K^0,$\\
    & $           $  & $D_1(2430)^0\pi^{0}, D_1(2430)^+\pi^{-}, D_1(2430)^0\eta,$            & $           $  & $D_1(2430)^0\pi^{+}, D_1(2430)^+\pi^{0}, D_1(2430)^+\eta,$\\
    & $           $  & $D_{s1}(2460)^{+}K^-$                                                 & $           $  & $D_{s1}(2460)^{+}K^0,$\\
    & $           $  & $D(2P_1)^0\pi^{0}, D(2P_1)^+\pi^{-}, $                                 & $           $  & $D(2P_1)^0\pi^{+}, D(2P_1)^+\pi^{0}$\\
    & $           $  & $D(2P_1^\prime)^0\pi^{0}, D(2P_1^\prime)^+\pi^{-} $                   & $           $  & $D(2P_1^\prime)^0\pi^{+}, D(2P_1^\prime)^+\pi^{0}$\\
                  & $ $  & $  $      & $           $  & $        $ \\

    & $1^{+}+1^{-}$  & $D_1(2420)^0\rho(770)^0, D_1(2420)^+\rho(770)^-,$                     & $1^{+}+1^{-}$  & $D_1(2420)^0\rho(770)^+, D_1(2420)^+\rho(770)^0,$ \\
    & $           $  & $D_1(2420)^0\omega,$                                                  & $           $  & $D_1(2420)^+\omega,$ \\
    & $           $  & $D_1(2430)^0\rho(770)^0, D_1(2430)^+\rho(770)^-, $                    & $           $  & $D_1(2430)^0\rho(770)^+, D_1(2430)^+\rho(770)^0,$ \\
    & $           $  & $D_1(2430)^0\omega $                                                  & $           $  & $D_1(2430)^+\omega $ \\
                  & $ $  & $  $      & $           $  & $        $ \\

    & $2^{+}+0^{-}$  &$D_2^*(2460)^0\pi^0, D_2^*(2460)^+\pi^-, D_{s2}^*(2573)^+K^-,$         & $2^{+}+0^{-}$  & $D_2^*(2460)^0\pi^+, D_2^*(2460)^+\pi^0, D_{s2}^*(2573)^+K^0,$\\
    & $           $  &$D(2^3P_2)^0\pi^0, D(2^3P_2)^+\pi^-$                                   &                & $D(2^3P_2)^0\pi^+, D(2^3P_2)^+\pi^0 $\\
                  & $ $  & $  $      & $           $  & $        $ \\


    & $2^{-}+0^{-}$  &$D(1^1D_2)\pi^0, (1^1D_2)D(2750)^+\pi^-, $                             & $2^{-}+0^{-}$  & $D(1^1D_2)\pi^+, (1^1D_2)D(2750)^+\pi^0,$\\
    & $           $  &$D(1^3D_2)\pi^0, (1^3D_2)D(2750)^+\pi^- $                              &                & $D(1^3D_2)\pi^+, (1^3D_2)D(2750)^+\pi^0 $\\
                  & $ $  & $  $      & $           $  & $        $ \\

    & $3^{-}+0^{-}$  &$D(1^3D_3)^0\pi^0, D(1^3D_3)^+\pi^-$                                   & $3^{-}+0^{-}$  & $D(1^3D_3)^0\pi^+, D(1^3D_3)^+\pi^0 $\\
                  & $ $  & $  $      & $           $  & $        $ \\
     \hline\hline
\end{tabular}
\label{table3}
\end{table*}
Numerical results of the decay width for all possible hadronic channels from the four radially excited $2D~D$ resonances are presented in from Table VI to Table XIII. As introduced in the first section, there are different assignments to $D(2550)$, $D^*(2600)$, $D(2750)$, $D^*_1(2760)$ and $D^*_3(2760)$. In order to present the decays to these final resonances explicitly, two tables for each radially excited $2D~D$ resonance are given. All the decay channels except for $D(2^3S_1)\pi$, $D(1^3D_1)\pi$, $D(1^3D_3)\pi$, $D(1^1D_2)\pi$ and $D(1^3D_2)\pi$ are presented in the first kind of tables (Table VI, VIII, X and XII), where the $\Gamma^\prime$ indicates the `total` decay width of all channels in each table.

Numerical results for hadronic channels such as $D(2^3S_1)\pi$, $D(1^3D_1)\pi$, $D(1^3D_3)\pi$, $D(1^1D_2)\pi$ and $D(1^3D_2)\pi$ in different assignments are given in the second kind tables (Table VII, IX, XI and XIII). These channels contribute to the total decay width $\Gamma$, which is a little larger than $\Gamma^\prime$.

For hadronic decays of $D(2^3D_1)$, the dominant decay channels are $D(2^3D_1)\to D^*\rho(770)$, $D(2^3D_1)\to D_1(2420)\pi$, $D(2^3D_1)\to D\pi(1300)$, $D(2^3D_1)\to D_1(2420)\rho(770)$, and $D(2^3D_1)\to D^*_2(2460)\pi$, etc.
We find the following ratios:
$$\frac{\Gamma_{D(2^3D_1)^0 \to D^{*0}\pi^0}}{\Gamma_{D(2^3D_1)^0 \to D^0\pi^0}}=0.46, \frac{\Gamma_{D(2^3D_1)^0 \to D^{*+}\pi^-}}{\Gamma_{D(2^3D_1)^0 \to D^+\pi^-}}=0.45,$$

$$ \frac{\Gamma_{D(2^3D_1)^0 \to D^{*0}\rho(770)^0}}{\Gamma_{D(2^3D_1)^0 \to D^*\pi}}=6.53, \frac{\Gamma_{D(2^3D_1)^0 \to D^{*+}\rho(770)^-}}{\Gamma_{D(2^3D_1)^0 \to D^*\pi}}=6.70.$$

\begin{table*}[t]
\caption{Hadronic decay widths of $D(2^3D_1)$ (in MeV).}
\begin{tabular}{p{0cm}p{3.4cm}p{1.2cm}p{3.2cm}p{3.4cm}p{1.2cm}p{2.2cm}}
   \hline\hline
  &$D(2^3D_1)$as $c\bar{u} $ &&& $D(2^3D_1)$ as $c\bar{d}$\\
   \hline

   & Channels                            & Width    & $ $   &Channels                              &   Width   & $     $   \\
   \hline
   &$D^0\pi^0$                           &  $3.7  $ & $ $   &$D^0\pi^+$                            &   $7.5 $   & $ $   \\  
   &$D^+\pi^-$                           &  $7.4  $ & $ $   &$D^+\pi^0$                            &   $3.7 $   & $ $   \\  
   &$D^0\eta $                           &  $1.3  $ & $ $   &$D^+\eta $                            &   $1.3 $   & $ $   \\  
   &$D^0\eta^\prime $                    &  $0.0  $ & $ $   &$D^+\eta^\prime $                     &   $0.0 $   & $ $   \\  
   &$D_s^+K^-$                           &  $2.9  $ & $ $   &$D_s^+K^0$                            &   $2.9 $   & $ $   \\  
   &$D^0\pi(1300)^0$                     &  $8.0  $ & $ $   &$D^0\pi(1300)^+$                      &   $16.0$   & $ $   \\  
   &$D^+\pi(1300)^-$                     &  $14.7 $ & $ $   &$D^+\pi(1300)^0$                      &   $7.4 $   & $ $   \\  
   &$D(2550)^0\pi^0$                     &  $1.2  $ & $ $   &$D(2550)^0\pi^+$                      &   $2.5 $   & $ $   \\  
   &$D(2550)^+\pi^-$                     &  $2.5  $ & $ $   &$D(2550)^+\pi^0$                      &   $1.2 $   & $ $   \\  
   &$D(2550)^0\eta $                     &  $3.0  $ & $ $   &$D(2550)^+\eta $                      &   $3.0 $   & $ $   \\  
   &$D_s(2^1S_0)^+K^-$                   &  $5.4  $ & $ $   &$D_s(2^1S_0)^+K^0$                    &   $5.0 $   & $ $   \\  

   &$D^0\rho(770)^0$                     &  $0.5  $ & $ $   &$D^0\rho(770)^+$                      &   $1.0 $   & $ $   \\  
   &$D^+\rho(770)^-$                     &  $1.0  $ & $ $   &$D^+\rho(770)^0$                      &   $0.5 $   & $ $   \\  
   &$D^0\phi(1020)$                      &  $0.0  $ & $ $   &$D^+\phi(1020)$                       &   $0.0 $   & $ $   \\  
   &$D^0\omega $                         &  $0.5  $ & $ $   &$D^+\omega $                          &   $0.5 $   & $ $   \\  
   &$D_s^+K^*(892)^-$                    &  $0.0  $ & $ $   &$D_s^+K^*(892)^0$                     &   $0.0 $   & $ $   \\  

   &$D^0b_1(1235)^0$                     &  $3.8  $ & $ $   &$D^0b_1(1235)^+ $                     &   $7.7 $   & $ $   \\  
   &$D^+b_1(1235)^-$                     &  $8.0  $ & $ $   &$D^+b_1(1235)^0$                      &   $4.0 $   & $ $   \\  
   &$D^0a_1(1260)^0$                     &  $1.7  $ & $ $   &$D^0a_1(1260)^+$                      &   $3.4 $   & $ $   \\  
   &$D^+a_1(1260)^-$                     &  $3.8  $ & $ $   &$D^+a_1(1260)^0$                      &   $1.9 $   & $ $   \\  

   &$D^0a_2(1320)^0$                     &  $0.2  $ & $ $   &$D^0a_2(1320)^+$                      &   $0.4 $   & $ $   \\  
   &$D^+a_2(1320)^-$                     &  $0.3  $ & $ $   &$D^+a_2(1320)^0$                      &   $0.5 $   & $ $   \\  

   &$D^{*0}\pi^0$                        &  $1.7  $ & $ $   &$D^{*0}\pi^+$                         &   $3.4 $   & $ $   \\  
   &$D^{*+}\pi^-$                        &  $3.3  $ & $ $   &$D^{*+}\pi^0$                         &   $1.7 $   & $ $   \\  
   &$D^{*0}\eta $                        &  $0.4  $ & $ $   &$D^{*+}\eta $                         &   $0.3 $   & $ $   \\  
   &$D^{*0}\eta^\prime $                 &  $0.2  $ & $ $   &$D^{*+}\eta^\prime $                  &   $0.3 $   & $ $   \\  
   &$D_s^{*+}K^-$                        &  $0.1  $ & $ $   &$D_s^{*+}K^0$                         &   $0.1 $   & $ $   \\  
   &$D_{s1}^*(2700)^+K^-$                &  $1.3  $ & $ $   &$D_{s1}^*(2700)^+K^0$                 &   $1.0 $   & $ $   \\  

   &$D^{*0}\rho(770)^0$                  &  $11.1 $ & $ $   &$D^{*0}\rho(770)^+$                   &   $22.3$   & $ $   \\  
   &$D^{*+}\rho(770)^-$                  &  $22.1 $ & $ $   &$D^{*+}\rho(770)^0$                   &   $11.0$   & $ $   \\  
   &$D^{*0}\phi(1020) $                  &  $3.8  $ & $ $   &$D^{*+}\phi(1020) $                   &   $3.8 $   & $ $   \\  
   &$D^{*0}\omega $                      &  $11.0 $ & $ $   &$D^{*+}\omega  $                      &   $10.9$   & $ $   \\  
   &$D_s^{*+}K^*(892)^-$                 &  $3.6  $ & $ $   &$D_s^{*+}K^*(892)^0$                  &   $3.6 $   & $ $   \\  

   &$D_0^*(2400)^0\rho(770)^0$           &  $1.1  $ & $ $   &$D_0^*(2400)^0\rho(770)^+$            &   $2.1 $   & $ $   \\  
   &$D_0^*(2400)^+\rho(770)^-$           &  $0.3  $ & $ $   &$D_0^*(2400)^+\rho(770)^0$            &   $0.1 $   & $ $   \\  
   &$D_0^*(2400)^0\omega   $             &  $1.0  $ & $ $   &$D_0^*(2400)^+\omega $                &   $0.1 $   & $ $   \\  
   &$D_{s0}^*(2317)^{+}K^*(892)^-$       &  $0.0  $ & $ $   &$D_{s0}^*(2317)^{+}K^*(892)^0$        &   $0.0 $   & $ $   \\  

   &$D_1(2420)^0\pi^0$                   &  $9.5  $ & $ $   &$D_1(2420)^0\pi^+$                    &   $18.9$   & $ $   \\  
   &$D_1(2420)^+\pi^-$                   &  $18.8 $ & $ $   &$D_1(2420)^+\pi^0$                    &   $9.4 $   & $ $   \\  
   &$D_1(2420)^0\eta $                   &  $2.1  $ & $ $   &$D_1(2420)^+\eta $                    &   $2.1 $   & $ $   \\  
   &$D_{s1}(2536)^+K^-$                  &  $10.4 $ & $ $   &$D_{s1}(2536)^+K^0$                   &   $10.6$   & $ $   \\  
   &$D_1(2430)^0\pi^0$                   &  $0.6  $ & $ $   &$D_1(2430)^0\pi^+$                    &   $1.2 $   & $ $   \\  
   &$D_1(2430)^+\pi^-$                   &  $1.2  $ & $ $   &$D_1(2430)^+\pi^0$                    &   $0.6 $   & $ $   \\  
   &$D_1(2430)^0\eta $                   &  $0.3  $ & $ $   &$D_1(2430)^0\eta $                    &   $0.3 $   & $ $   \\  
   &$D_{s1}(2460)^{+}K^-$                &  $0.8  $ & $ $   &$D_{s1}(2460)^{+}K^0$                 &   $0.9 $   & $ $   \\  
   &$D(2P_1)^0\pi^0$                     &  $6.0  $ & $ $   &$D(2P_1)^0\pi^+$                      &   $12.1$   & $ $   \\  
   &$D(2P_1)^+\pi^-$                     &  $12.1 $ & $ $   &$D(2P_1)^+\pi^0$                      &   $6.0 $   & $ $   \\  
   &$D(2P_1^\prime)^0\pi^0$              &  $0.1  $ & $ $   &$D(2P_1^\prime)^0\pi^+$               &   $0.1 $   & $ $   \\  
   &$D(2P_1^\prime)^+\pi^-$              &  $0.1  $ & $ $   &$D(2P_1^\prime)^+\pi^0$               &   $0.1 $   & $ $   \\  

   &$D_1(2420)^0\rho(770)^0$             &  $7.1  $ & $ $   &$D_1(2420)^0\rho(770)^+$              &   $14.2$   & $ $   \\  
   &$D_1(2420)^+\rho(770)^-$             &  $14.1 $ & $ $   &$D_1(2420)^+\rho(770)^0$              &   $7.1 $   & $ $   \\  
   &$D_1(2420)^0\omega $                 &  $7.0  $ & $ $   &$D_1(2420)^+\omega $                  &   $6.9 $   & $ $   \\  
   &$D_1(2430)^0\rho(770)^0$             &  $0.3  $ & $ $   &$D_1(2430)^0\rho(770)^+$              &   $0.6 $   & $ $   \\  
   &$D_1(2430)^+\rho(770)^-$             &  $0.6  $ & $ $   &$D_1(2430)^+\rho(770)^0$              &   $0.3 $   & $ $   \\  
   &$D_1(2430)^0\omega $                 &  $0.2  $ & $ $   &$D_1(2430)^+\omega $                  &   $0.2 $   & $ $   \\  

   &$D_2^*(2460)^0\pi^0$                 &  $5.1  $ & $ $   &$D_2^*(2460)^0\pi^+$                  &   $10.2$   & $ $   \\  
   &$D_2^*(2460)^+\pi^-$                 &  $10.2 $ & $ $   &$D_2^*(2460)^+\pi^0$                  &   $5.1 $   & $ $   \\  
   &$D_2^*(2460)^0\eta $                 &  $1.7  $ & $ $   &$D_2^*(2460)^+\eta $                  &   $1.7 $   & $ $   \\  
   &$D_{s2}^*(2573)^+K^-$                &  $2.2  $ & $ $   &$D_{s2}^*(2573)^+K^0$                 &   $2.1 $   & $ $   \\  
   &$D(2^3P_2)^0\pi^0$                   &  $0.2  $ & $ $   &$D(2^3P_2)^0\pi^+$                    &   $0.3 $   & $ $   \\  
   &$D(2^3P_2)^+\pi^-$                   &  $0.3  $ & $ $   &$D(2^3P_2)^+\pi^0$                    &   $0.2 $   & $ $   \\  

   &  $\Gamma^\prime$                   &  $246.4$ & $ $   & $\Gamma^\prime$                      &   $246.2$  & $ $   \\
         \hline\hline
\end{tabular}
\label{table5}
\end{table*}

\begin{table*}[t]
\caption{Hadronic decay widths of $D(2^3D_1)$ (in MeV).}
\begin{tabular}{p{0cm}p{3.4cm}p{1.2cm}p{3.2cm}p{3.4cm}p{1.2cm}p{2.2cm}}
   \hline\hline
  &$D(2^3D_1)$as $c\bar{u} $ &&& $D(2^3D_1)$ as $c\bar{d}$\\
   \hline

   & Channels                            & Width    & $ $   &Channels                              &  Width     & $     $   \\
   \hline
   &$D^*(2600)^0(2^3S_1)\pi^0$             &  $1.3  $ & $ $   &$D^*(2600)^0(2^3S_1)\pi^+$              &   $2.7 $   & $ $   \\  
   &$D^*(2600)^+(2^3S_1)\pi^-$             &  $2.5  $ & $ $   &$D^*(2600)^+(2^3S_1)\pi^0$              &   $1.2 $   & $ $   \\  
   &$D^*(2600)^0(1^3D_1)\pi^0$             &  $0.0  $ & $ $   &$D^*(2600)^0(1^3D_1)\pi^+$              &   $0.0 $   & $ $   \\  
   &$D^*(2600)^+(1^3D_1)\pi^-$             &  $0.0  $ & $ $   &$D^*(2600)^+(1^3D_1)\pi^0$              &   $0.0 $   & $ $   \\  
   &$                      $             &  $     $ & $     $   &$                      $              &   $    $   & $     $   \\

   &$D(2750)^0(1^1D_2)\pi^0$             &  $12.4 $ & $ $   &$D(2750)^0(1^1D_2)\pi^+$              &   $25.0$   & $ $   \\  
   &$D(2750)^0(1^3D_2)\pi^0$             &  $8.6  $ & $ $   &$D(2750)^0(1^3D_2)\pi^+$              &   $17.2$   & $ $   \\  
   &$                      $             &  $     $ & $     $   &$                      $              &   $    $   & $     $   \\


   &$D_1^*(2760)^0(1^3D_1)\pi^0$         &  $0.0  $ & $ $   &$D_1^*(2760)^0(1^3D_1)\pi^+$          &   $0.0 $   & $ $   \\  
   &$D_1^*(2760)^+(1^3D_1)\pi^-$         &  $0.0  $ & $ $   &$D_1^*(2760)^+(1^3D_1)\pi^0$          &   $0.0 $   & $ $   \\  
   &$D_3^*(2760)^0(1^3D_3)\pi^0$         &  $0.3  $ & $ $   &$D_3^*(2760)^0(1^3D_3)\pi^+$          &   $0.7 $   & $ $   \\  
   &$D_3^*(2760)^+(1^3D_3)\pi^-$         &  $0.6  $ & $ $   &$D_3^*(2760)^+(1^3D_3)\pi^0$          &   $0.3 $   & $ $   \\  

         \hline\hline
\end{tabular}
\label{table5}
\end{table*}
For hadronic decays of $D(2^3D_3)$, the dominant decay channels are $D(2^3D_3)\to D^*_2(2460)\pi$, $D(2^3D_3)\to D^*\rho(770)$, $D(2^3D_3)\to D_1(2420)\pi$, and $D(2^3D_3)\to D(2550)\pi$, etc. We find the following ratios:
$$ \frac{\Gamma_{D(2^3D_3)^0 \to D^{*0}\rho(770)^0}}{\Gamma_{D(2^3D_3)^0 \to D^{*0}\pi^0}}=7.40 , \frac{\Gamma_{D(2^3D_3)^0 \to D^{*+}\rho(770)^-}}{\Gamma_{D(2^3D_3)^0 \to D^{*+}\pi^-}}=6.55 .$$
\begin{table*}[t]
\caption{Hadronic decay widths of $D(2^3D_3)$ (in MeV).}
\begin{tabular}{p{0cm}p{3.4cm}p{1.2cm}p{3.2cm}p{3.4cm}p{1.2cm}p{2.2cm}}
   \hline\hline
  &$D(2^3D_3)$as $c\bar{u} $ &&& $D(2^3D_3)$ as $c\bar{d}$\\
   \hline

   & Channels                            & Width    & $  $   &Channels                              &  Width   & $     $   \\
     \hline
   &$D^0\pi^0$                           &  $0.0  $ & $  $   &$D^0\pi^+$                            &   $0.0 $   & $  $   \\  
   &$D^+\pi^-$                           &  $0.0  $ & $  $   &$D^+\pi^0$                            &   $0.0 $   & $  $   \\  
   &$D^0\eta $                           &  $0.1  $ & $  $   &$D^+\eta $                            &   $0.1 $   & $  $   \\  
   &$D^0\eta^\prime $                    &  $0.3  $ & $  $   &$D^+\eta^\prime $                     &   $0.3 $   & $  $   \\  
   &$D_s^+K^-$                           &  $0.4  $ & $  $   &$D_s^+K^0$                            &   $0.4 $   & $  $   \\  
   &$D^0\pi(1300)^0$                     &  $0.1  $ & $  $   &$D^0\pi(1300)^+$                      &   $0.1 $   & $  $   \\  
   &$D^+\pi(1300)^-$                     &  $0.1  $ & $  $   &$D^+\pi(1300)^0$                      &   $0.0 $   & $  $   \\  
   &$D(2550)^0\pi^0$                     &  $3.0  $ & $  $   &$D(2550)^0\pi^+$                      &   $6.1 $   & $  $   \\  
   &$D(2550)^+\pi^-$                     &  $6.1  $ & $  $   &$D(2550)^+\pi^0$                      &   $3.0 $   & $  $   \\  
   &$D(2550)^0\eta $                     &  $0.2  $ & $  $   &$D(2550)^+\eta $                      &   $0.2 $   & $  $   \\  
   &$D_s(2^1S_0)^+K^-$                   &  $0.0  $ & $  $   &$D_s(2^1S_0)^+K^0$                    &   $0.0 $   & $  $   \\  

   &$D^0\rho(770)^0$                     &  $0.6  $ & $  $   &$D^0\rho(770)^+$                      &   $1.3 $   & $  $   \\  
   &$D^+\rho(770)^-$                     &  $1.3  $ & $  $   &$D^+\rho(770)^0$                      &   $0.6 $   & $  $   \\  
   &$D^0\phi(1020)$                      &  $0.6  $ & $  $   &$D^+\phi(1020)$                       &   $0.6 $   & $  $   \\  
   &$D^0\omega $                         &  $0.6  $ & $  $   &$D^+\omega $                          &   $0.6 $   & $  $   \\  
   &$D_s^+K^*(892)^-$                    &  $0.8  $ & $  $   &$D_s^+K^*(892)^0$                     &   $0.8 $   & $  $   \\  

   &$D^0b_1(1235)^0$                     &  $1.5  $ & $  $   &$D^0b_1(1235)^+ $                     &   $3.0 $   & $  $   \\  
   &$D^+b_1(1235)^-$                     &  $2.8  $ & $  $   &$D^+b_1(1235)^0$                      &   $1.4 $   & $  $   \\  
   &$D^0a_1(1260)^0$                     &  $0.7  $ & $  $   &$D^0a_1(1260)^+$                      &   $1.3 $   & $  $   \\  
   &$D^+a_1(1260)^-$                     &  $1.2  $ & $  $   &$D^+a_1(1260)^0$                      &   $0.6 $   & $  $   \\  

   &$D^0a_2(1320)^0$                     &  $0.1  $ & $  $   &$D^0a_2(1320)^+$                      &   $0.3 $   & $  $   \\  
   &$D^+a_2(1320)^-$                     &  $0.2  $ & $  $   &$D^+a_2(1320)^0$                      &   $0.3 $   & $  $   \\  

   &$D^{*0}\pi^0$                        &  $0.5  $ & $  $   &$D^{*0}\pi^+$                         &   $1.1 $   & $  $   \\  
   &$D^{*+}\pi^-$                        &  $1.1  $ & $  $   &$D^{*+}\pi^0$                         &   $0.6 $   & $  $   \\  
   &$D^{*0}\eta $                        &  $0.7  $ & $  $   &$D^{*+}\eta $                         &   $0.7 $   & $  $   \\  
   &$D^{*0}\eta^\prime $                 &  $0.4  $ & $  $   &$D^{*+}\eta^\prime $                  &   $0.4 $   & $  $   \\  
   &$D_s^{*+}K^-$                        &  $1.8  $ & $  $   &$D_s^{*+}K^0$                         &   $1.8 $   & $  $   \\  
   &$D_{s1}^*(2700)^+K^-$                &  $0.0  $ & $  $   &$D_{s1}^*(2700)^+K^0$                 &   $0.0 $   & $  $   \\  

   &$D^{*0}\rho(770)^0$                  &  $3.7  $ & $  $   &$D^{*0}\rho(770)^+$                   &   $7.4 $   & $  $   \\  
   &$D^{*+}\rho(770)^-$                  &  $7.2  $ & $  $   &$D^{*+}\rho(770)^0$                   &   $3.6 $   & $  $   \\  
   &$D^{*0}\phi(1020) $                  &  $2.1  $ & $  $   &$D^{*+}\phi(1020) $                   &   $2.2 $   & $  $   \\  
   &$D^{*0}\omega $                      &  $3.5  $ & $  $   &$D^{*+}\omega  $                      &   $3.4 $   & $  $   \\  
   &$D_s^{*+}K^*(892)^-$                 &  $3.6  $ & $  $   &$D_s^{*+}K^*(892)^0$                  &   $3.6 $   & $  $   \\  



   &$D_0^*(2400)^0\rho(770)^0$           &  $0.7  $ & $  $   &$D_0^*(2400)^0\rho(770)^+$            &   $1.5 $   & $  $   \\  
   &$D_0^*(2400)^+\rho(770)^-$           &  $0.2  $ & $  $   &$D_0^*(2400)^+\rho(770)^0$            &   $0.1 $   & $  $   \\  
   &$D_0^*(2400)^0\omega   $             &  $0.7  $ & $  $   &$D_0^*(2400)^+\omega $                &   $0.1 $   & $  $   \\  
   &$D_{s0}^*(2317)^{+}K^*(892)^-$       &  $0.0  $ & $  $   &$D_{s0}^*(2317)^{+}K^*(892)^0$        &   $0.0 $   & $  $   \\  

   &$D_1(2420)^0\pi^0$                   &  $3.6  $ & $  $   &$D_1(2420)^0\pi^+$                    &   $7.1 $   & $  $   \\  
   &$D_1(2420)^+\pi^-$                   &  $7.1  $ & $  $   &$D_1(2420)^+\pi^0$                    &   $3.5 $   & $  $   \\  
   &$D_1(2420)^0\eta $                   &  $0.5  $ & $  $   &$D_1(2420)^+\eta $                    &   $0.5 $   & $  $   \\  
   &$D_{s1}(2536)^+K^-$                  &  $1.6  $ & $  $   &$D_{s1}(2536)^+K^0$                   &   $1.6 $   & $  $   \\  
   &$D_1(2430)^0\pi^0$                   &  $1.7  $ & $  $   &$D_1(2430)^0\pi^+$                    &   $3.4 $   & $  $   \\  
   &$D_1(2430)^+\pi^-$                   &  $3.4  $ & $  $   &$D_1(2430)^+\pi^0$                    &   $1.7 $   & $  $   \\  
   &$D_1(2430)^0\eta $                   &  $0.8  $ & $  $   &$D_1(2430)^0\eta $                    &   $0.8 $   & $  $   \\  
   &$D_{s1}(2460)^{+}K^-$                &  $1.2  $ & $  $   &$D_{s1}(2460)^{+}K^0$                 &   $1.2 $   & $  $   \\  
   &$D(2P_1)^0\pi^0$                     &  $0.1  $ & $  $   &$D(2P_1)^0\pi^+$                      &   $0.1 $   & $ $   \\  
   &$D(2P_1)^+\pi^-$                     &  $0.1  $ & $  $   &$D(2P_1)^+\pi^0$                      &   $0.0 $   & $ $   \\  
   &$D(2P_1^\prime)^0\pi^0$              &  $0.0  $ & $  $   &$D(2P_1^\prime)^0\pi^+$               &   $0.1 $   & $ $   \\  
   &$D(2P_1^\prime)^+\pi^-$              &  $0.1  $ & $  $   &$D(2P_1^\prime)^+\pi^0$               &   $0.1 $   & $ $   \\  

   &$D_1(2420)^0\rho(770)^0$             &  $0.1  $ & $  $   &$D_1(2420)^0\rho(770)^+$              &   $0.1 $   & $  $   \\  
   &$D_1(2420)^+\rho(770)^-$             &  $0.1  $ & $  $   &$D_1(2420)^+\rho(770)^0$              &   $0.0 $   & $  $   \\  
   &$D_1(2420)^0\omega $                 &  $0.0  $ & $  $   &$D_1(2420)^+\omega $                  &   $0.0 $   & $  $   \\  
   &$D_1(2430)^0\rho(770)^0$             &  $0.1  $ & $  $   &$D_1(2430)^0\rho(770)^+$              &   $0.1 $   & $  $   \\  
   &$D_1(2430)^+\rho(770)^-$             &  $0.1  $ & $  $   &$D_1(2430)^+\rho(770)^0$              &   $0.1 $   & $  $   \\  
   &$D_1(2430)^0\omega $                 &  $0.0  $ & $  $   &$D_1(2430)^+\omega $                  &   $0.0 $   & $  $   \\  

   &$D_2^*(2460)^0\pi^0$                 &  $3.7  $ & $  $   &$D_2^*(2460)^0\pi^+$                  &   $7.4 $   & $  $   \\  
   &$D_2^*(2460)^+\pi^-$                 &  $7.4  $ & $  $   &$D_2^*(2460)^+\pi^0$                  &   $3.7 $   & $  $   \\  
   &$D_2^*(2460)^0\eta $                 &  $0.9  $ & $  $   &$D_2^*(2460)^+\eta $                  &   $0.9 $   & $  $   \\  
   &$D_{s2}^*(2573)^+K^-$                &  $1.2  $ & $  $   &$D_{s2}^*(2573)^+K^0$                 &   $1.2 $   & $  $   \\  
   &$D(2^3P_2)^0\pi^0$                   &  $0.1  $ & $  $   &$D(2^3P_2)^0\pi^+$                    &   $0.2 $   & $  $   \\  
   &$D(2^3P_2)^+\pi^-$                   &  $0.2  $ & $  $   &$D(2^3P_2)^+\pi^0$                    &   $0.1 $   & $  $   \\  


   &$\Gamma^\prime$                      &  $87.9 $ & $  $   & $\Gamma^\prime$                       &   $88.1 $  & $ $   \\

         \hline\hline
\end{tabular}
\label{table5}
\end{table*}

\begin{table*}[t]
\caption{Hadronic decay widths of $D(2^3D_3)$ (in MeV).}
\begin{tabular}{p{0cm}p{3.4cm}p{1.2cm}p{3.2cm}p{3.4cm}p{1.2cm}p{2.2cm}}
   \hline\hline
  &$D(2^3D_1)$as $c\bar{u} $ &&& $D(2^3D_3)$ as $c\bar{d}$\\
   \hline

   & Channels                            & Width    & $     $   &Channels                              &   Width    & $     $   \\
   \hline
   &$D^*(2600)^0(2^3S_1)\pi^0$             &  $2.5  $ & $  $   &$D^*(2600)^0(2^3S_1)\pi^+$              &   $5.1 $   & $  $   \\  
   &$D^*(2600)^+(2^3S_1)\pi^-$             &  $5.4  $ & $  $   &$D^*(2600)^+(2^3S_1)\pi^0$              &   $2.7 $   & $  $   \\  
   &$D^*(2600)^0(1^3D_1)\pi^0$             &  $0.1  $ & $  $   &$D^*(2600)^0(1^3D_1)\pi^+$              &   $0.1 $   & $  $   \\  
   &$D^*(2600)^+(1^3D_1)\pi^-$             &  $0.0  $ & $  $   &$D^*(2600)^+(1^3D_1)\pi^0$              &   $0.0 $   & $  $   \\  
   &$                      $             &  $     $ & $     $   &$                      $              &   $    $   & $     $   \\

   &$D(2750)^0(1^1D_2)\pi^0$             &  $0.6  $ & $  $   &$D(2750)^0(1^1D_2)\pi^+$              &   $1.1 $   & $  $   \\  
   &$D(2750)^0(1^3D_2)\pi^0$             &  $0.1  $ & $  $   &$D(2750)^0(1^3D_2)\pi^+$              &   $0.3 $   & $  $   \\  
   &$                      $             &  $     $ & $     $   &$                      $              &   $    $   & $     $   \\


   &$D_1^*(2760)^0(1^3D_1)\pi^0$         &  $0.0  $ & $ $   &$D_1^*(2760)^0(1^3D_1)\pi^+$          &   $0.0 $   & $ $   \\  
   &$D_1^*(2760)^+(1^3D_1)\pi^-$         &  $0.0  $ & $ $   &$D_1^*(2760)^+(1^3D_1)\pi^0$          &   $0.0 $   & $ $   \\  
   &$D_3^*(2760)^0(1^3D_3)\pi^0$         &  $0.9  $ & $ $   &$D_3^*(2760)^0(1^3D_3)\pi^+$          &   $1.8 $   & $ $   \\  
   &$D_3^*(2760)^+(1^3D_3)\pi^-$         &  $1.8  $ & $ $   &$D_3^*(2760)^+(1^3D_3)\pi^0$          &   $0.9 $   & $ $   \\  
         \hline\hline
\end{tabular}
\label{table5}
\end{table*}
For hadronic decays of $D(2D_2)$, the dominant decay channels are $D(2D_2)\to D_1(2420)\rho(770)$, $D(2D_2)\to Da_2(1320)$, $D(2D_2)\to D^*_2(2460)\pi$, and $D(2D_2)\to D^*\rho(770)$, etc. We find the following ratios:
$$\frac{\Gamma_{D(2D_2)^0 \to D^{*0}\rho(770)^0}}{\Gamma_{D(2D_2)^0 \to D^0\rho(770)^0}}=2.24, \frac{\Gamma_{D(2D_2)^0 \to D^{*+}\rho(770)^-}}{\Gamma_{D(2D_2)^0 \to D^+\rho(770)^-}}=2.27,$$
$$\frac{\Gamma_{D(2D_2)^0 \to D^{*0}\pi^0}}{\Gamma_{D(2D_2)^0 \to D^0\rho(770)^0}}=1.05, \frac{\Gamma_{D(2D_2)^0 \to D^{*+}\pi^-}}{\Gamma_{D(2D_2)^0 \to D^+\rho(770)^-}}=1.10.$$
\begin{table*}[t]
\caption{Hadronic decay widths of $D(2 D_2)$ (in MeV).}
\begin{tabular}{p{0cm}p{3.4cm}p{1.2cm}p{3.2cm}p{3.4cm}p{1.2cm}p{2.2cm}}
   \hline\hline
  &$D(2 D_2)$as $c\bar{u} $ &&& $D(2 D_2)$ as $c\bar{d}$\\
   \hline

   & Channels                            & Width    & $     $   &Channels                           &   Width    & $     $   \\
   \hline
   &$D^0\rho(770)^0$                     &  $2.1  $ & $  $   &$D^0\rho(770)^+$                      &   $4.2 $   & $  $   \\  
   &$D^+\rho(770)^-$                     &  $4.1  $ & $  $   &$D^+\rho(770)^0$                      &   $2.0 $   & $  $   \\  
   &$D^0\phi(1020) $                     &  $0.1  $ & $  $   &$D^+\phi(1020) $                      &   $0.0 $   & $  $   \\  
   &$D^0\omega  $                        &  $2.0  $ & $  $   &$D^+\omega  $                         &   $2.0 $   & $  $   \\  
   &$D_s^+K^*(892)^-$                    &  $0.6  $ & $  $   &$D_s^+K^*(892)^0$                     &   $0.6 $   & $  $   \\  

   &$D^0b_1(1235)^0$                     &  $0.1  $ & $  $   &$D^0b_1(1235)^+ $                     &   $0.2 $   & $  $   \\  
   &$D^+b_1(1235)^-$                     &  $0.1  $ & $  $   &$D^+b_1(1235)^0$                      &   $0.1 $   & $  $   \\  
   &$D^0a_1(1260)^0$                     &  $0.3  $ & $  $   &$D^0a_1(1260)^+$                      &   $0.6 $   & $  $   \\  
   &$D^+a_1(1260)^-$                     &  $0.5  $ & $  $   &$D^+a_1(1260)^0$                      &   $0.3 $   & $  $   \\  

   &$D^0a_2(1320)^0$                     &  $10.1 $ & $  $   &$D^0a_2(1320)^+$                      &   $20.2$   & $  $   \\  
   &$D^+a_2(1320)^-$                     &  $20.2 $ & $  $   &$D^+a_2(1320)^0$                      &   $10.1$   & $  $   \\  

   &$D^{*0}\pi^0$                        &  $2.2  $ & $  $   &$D^{*0}\pi^+$                         &   $4.4 $   & $  $   \\  
   &$D^{*+}\pi^-$                        &  $4.5  $ & $  $   &$D^{*+}\pi^0$                         &   $2.2 $   & $  $   \\  
   &$D^{*0}\eta $                        &  $1.8  $ & $  $   &$D^{*+}\eta $                         &   $1.8 $   & $  $   \\  
   &$D^{*0}\eta^\prime $                 &  $0.8  $ & $  $   &$D^{*+}\eta^\prime $                  &   $0.8 $   & $  $   \\  
   &$D_s^{*+}K^-$                        &  $4.0  $ & $  $   &$D_s^{*+}K^0$                         &   $4.0 $   & $  $   \\  
   &$D_{s1}^*(2700)^+K^-$                &  $0.5  $ & $  $   &$D_{s1}^*(2700)^+K^0$                 &   $0.2 $   & $  $   \\  

   &$D^{*0}\rho(770)^0$                  &  $4.7  $ & $  $   &$D^{*0}\rho(770)^+$                   &   $9.5 $   & $  $   \\  
   &$D^{*+}\rho(770)^-$                  &  $9.3  $ & $  $   &$D^{*+}\rho(770)^0$                   &   $4.7 $   & $  $   \\  
   &$D^{*0}\phi(1020) $                  &  $2.4  $ & $  $   &$D^{*+}\phi(1020) $                   &   $2.4 $   & $  $   \\  
   &$D^{*0}\omega $                      &  $4.6  $ & $  $   &$D^{*+}\omega  $                      &   $4.6 $   & $  $   \\  
   &$D_s^{*+}K^*(892)^-$                 &  $2.9  $ & $  $   &$D_s^{*+}K^*(892)^0$                  &   $2.9 $   & $  $   \\  


   &$D_0^*(2400)^0\pi^0$                 &  $0.8  $ & $  $   &$D_0^*(2400)^0\pi^+$                  &   $1.7 $   & $  $   \\  
   &$D_0^*(2400)^+\pi^-$                 &  $2.0  $ & $  $   &$D_0^*(2400)^+\pi^0$                  &   $1.0 $   & $  $   \\  
   &$D_0^*(2400)^0\eta $                 &  $0.6  $ & $  $   &$D_0^*(2400)^+\eta $                  &   $0.5 $   & $  $   \\  
   &$D_{s0}^*(2317)^{+}K^-$              &  $2.2  $ & $  $   &$D_{s0}^*(2317)^{+}K^0$               &   $2.2 $   & $  $   \\  
   &$D(2^3P_0)^0\pi^0$                   &  $0.0  $ & $  $   &$D(2^3P_0)^0\pi^+$                    &   $0.1 $   & $  $   \\  
   &$D(2^3P_0)^+\pi^-$                   &  $0.1  $ & $  $   &$D(2^3P_0)^+\pi^0$                    &   $0.0 $   & $  $   \\  

   &$D_0^*(2400)^0\rho(770)^0$           &  $0.3  $ & $  $   &$D_0^*(2400)^0\rho(770)^+$            &   $0.7 $   & $  $   \\  
   &$D_0^*(2400)^+\rho(770)^-$           &  $0.0  $ & $  $   &$D_0^*(2400)^+\rho(770)^0$            &   $0.0 $   & $  $   \\  
   &$D_0^*(2400)^0\omega $               &  $0.3  $ & $  $   &$D_0^*(2400)^+\omega $                &   $0.0 $   & $  $   \\  
   &$D_{s0}^*(2317)^{+}K^*(892)^-$       &  $0.0  $ & $  $   &$D_{s0}^*(2317)^{+}K^*(892)^0      $  &   $--- $   & $  $   \\  

   &$D_1(2420)^0\pi^0$                   &  $1.1  $ & $  $   &$D_1(2420)^0\pi^+$                    &   $2.2 $   & $  $   \\  
   &$D_1(2420)^+\pi^-$                   &  $2.2  $ & $  $   &$D_1(2420)^+\pi^0$                    &   $1.1 $   & $  $   \\  
   &$D_1(2420)^0\eta$                    &  $0.5  $ & $  $   &$D_1(2420)^+\eta$                     &   $0.5 $   & $  $   \\  
   &$D_{s1}(2536)^+K^-$                  &  $0.2  $ & $  $   &$D_{s1}(2536)^+K^0$                   &   $0.2 $   & $  $   \\  
   &$D_1(2430)^0\pi^0$                   &  $1.1  $ & $  $   &$D_1(2430)^0\pi^+$                    &   $2.3 $   & $  $   \\  
   &$D_1(2430)^+\pi^-$                   &  $2.3  $ & $  $   &$D_1(2430)^+\pi^0$                    &   $1.1 $   & $  $   \\  
   &$D_1(2430)^0\eta$                    &  $0.5  $ & $  $   &$D_1(2430)^+\eta$                     &   $0.5 $   & $  $   \\  
   &$D_{s1}(2460)^{+}K^-$                &  $2.8  $ & $  $   &$D_{s1}(2460)^{+}K^0$                 &   $2.8 $   & $  $   \\  
   &$D(2P_1)^0\pi^0$                     &  $0.1  $ & $  $   &$D(2P_1)^0\pi^+$                      &   $0.3 $   & $ $   \\  
   &$D(2P_1)^+\pi^-$                     &  $0.3  $ & $  $   &$D(2P_1)^+\pi^0$                      &   $0.1 $   & $ $   \\  
   &$D(2P_1^\prime)^0\pi^0$              &  $0.0  $ & $  $   &$D(2P_1^\prime)^0\pi^+$               &   $0.0 $   & $ $   \\  
   &$D(2P_1^\prime)^+\pi^-$              &  $0.0  $ & $  $   &$D(2P_1^\prime)^+\pi^0$               &   $0.0 $   & $ $   \\  

   &$D_1(2420)^0\rho(770)^0$             &  $14.4 $ & $  $   &$D_1(2420)^0\rho(770)^+$              &   $28.9$   & $  $   \\  
   &$D_1(2420)^+\rho(770)^-$             &  $27.9 $ & $  $   &$D_1(2420)^+\rho(770)^0$              &   $13.9$   & $  $   \\  
   &$D_1(2420)^0\omega$                  &  $11.7 $ & $  $   &$D_1(2420)^+\omega$                   &   $10.5$   & $  $   \\  
   &$D_1(2430)^0\rho(770)^0$             &  $0.4  $ & $  $   &$D_1(2430)^0\rho(770)^+$              &   $0.7 $   & $  $   \\  
   &$D_1(2430)^+\rho(770)^-$             &  $0.7  $ & $  $   &$D_1(2430)^+\rho(770)^0$              &   $0.4 $   & $  $   \\  
   &$D_1(2430)^0\omega$                  &  $0.2  $ & $  $   &$D_1(2430)^+\omega$                   &   $0.2 $   & $  $   \\  

   &$D_2^*(2460)^0\pi^0$                 &  $6.0  $ & $  $   &$D_2^*(2460)^0\pi^+$                  &   $11.9$   & $  $   \\  
   &$D_2^*(2460)^+\pi^-$                 &  $11.8 $ & $  $   &$D_2^*(2460)^+\pi^0$                  &   $5.9 $   & $  $   \\  
   &$D_2^*(2460)^0\eta $                 &  $0.5  $ & $  $   &$D_2^*(2460)^+\eta $                  &   $0.5 $   & $  $   \\  
   &$D_{s2}^*(2573)^+K^-$                &  $0.4  $ & $  $   &$D_{s2}^*(2573)^+K^0$                 &   $0.4 $   & $  $   \\  
   &$D(2^3P_2)^0\pi^0$                   &  $0.0  $ & $  $   &$D(2^3P_2)^0\pi^+$                    &   $0.1 $   & $  $   \\  
   &$D(2^3P_2)^+\pi^-$                   &  $0.1  $ & $  $   &$D(2^3P_2)^+\pi^0$                    &   $0.0 $   & $  $   \\  

   &$\Gamma^\prime$                      &  $179.9$ & $  $   & $\Gamma^\prime$                       &   $178.7$  & $  $   \\

         \hline\hline
\end{tabular}
\label{table5}
\end{table*}

\begin{table*}[t]
\caption{Hadronic decay widths of $D(2 D_2)$ (in MeV).}
\begin{tabular}{p{0cm}p{3.4cm}p{1.2cm}p{3.2cm}p{3.4cm}p{1.2cm}p{2.2cm}}
   \hline\hline
  &$D(2 D_2)$as $c\bar{u} $ &&& $D(2 D_2)$ as $c\bar{d}$\\
   \hline

   & Channels                            & Width    & $  $   &Channels                              &   Width    & $     $   \\
   \hline
   &$D^*(2600)^0(2^3S_1)\pi^0$             &  $3.9  $ & $  $   &$D^*(2600)^0(2^3S_1)\pi^+$              &   $7.6 $   & $  $   \\  
   &$D^*(2600)^+(2^3S_1)\pi^-$             &  $8.1  $ & $  $   &$D^*(2600)^+(2^3S_1)\pi^0$              &   $4.1 $   & $  $   \\  
   &$D^*(2600)^0(1^3D_1)\pi^0$             &  $0.2  $ & $  $   &$D^*(2600)^0(1^3D_1)\pi^+$              &   $0.4 $   & $  $   \\  
   &$D^*(2600)^+(1^3D_1)\pi^-$             &  $0.4  $ & $  $   &$D^*(2600)^+(1^3D_1)\pi^0$              &   $0.2 $   & $  $   \\  
   &$                      $             &  $     $ & $     $   &$                      $              &   $    $   & $     $   \\


   &$D^*(2760)^0(1^3D_1)\pi^0$             &  $0.3  $ & $  $   &$D^*(2760)^0(1^3D_1)\pi^+$              &   $0.6 $   & $  $   \\  
   &$D^*(2760)^+(1^3D_1)\pi^-$             &  $0.6  $ & $  $   &$D^*(2760)^+(1^3D_1)\pi^0$              &   $0.3 $   & $  $   \\  
   &$                      $             &  $     $ & $     $   &$                      $              &   $    $   & $     $   \\

   &$D_1^*(2760)^0(1^3D_1)\pi^0$         &  $0.3  $ & $  $   &$D_1^*(2760)^0(1^3D_1)\pi^+$          &   $0.6 $   & $ $   \\  
   &$D_1^*(2760)^+(1^3D_1)\pi^-$         &  $0.6  $ & $  $   &$D_1^*(2760)^+(1^3D_1)\pi^0$          &   $0.3 $   & $ $   \\  
   &$D_3^*(2760)^0(1^3D_3)\pi^0$         &  $0.2  $ & $  $   &$D_3^*(2760)^0(1^3D_3)\pi^+$          &   $0.3 $   & $ $   \\  
   &$D_3^*(2760)^+(1^3D_3)\pi^-$         &  $0.3  $ & $  $   &$D_3^*(2760)^+(1^3D_3)\pi^0$          &   $0.2 $   & $ $   \\  

         \hline\hline
\end{tabular}
\label{table5}
\end{table*}
For hadronic decays of $D(2D_2^\prime)$, the dominant decay channels are $D(2D_2^\prime)\to D^*_2(2460)\pi$, $D(2D_2^\prime)\to D(2^3P_2)\pi$, $D(2D_2^\prime)\to D^*\pi$, and $D(2D_2^\prime)\to D^*\rho(770)$, etc.
We find the following ratios:
$$\frac{\Gamma_{D(2D_2^\prime)^0 \to D^{*0}\rho(770)^0}}{\Gamma_{D(2D_2^\prime)^0 \to D^0\rho(770)^0}}=3.17, \frac{\Gamma_{D(2D_2^\prime)^0 \to D^{*+}\rho(770)^-}}{\Gamma_{D(2D_2^\prime)^0 \to D^+\rho(770)^-}}=3.03,$$
$$\frac{\Gamma_{D(2D_2^\prime)^0 \to D^{*0}\pi^0}}{\Gamma_{D(2D_2^\prime)^0 \to D^0\rho(770)^0}}=3.56, \frac{\Gamma_{D(2D_2^\prime)^0 \to D^{*+}\pi^-}}{\Gamma_{D(2D_2^\prime)^0 \to D^+\rho(770)^-}}=3.41.$$
\begin{table*}[t]
\caption{Hadronic decay widths of $D(2D_2^\prime)$ (in MeV).}
\begin{tabular}{p{0cm}p{3.4cm}p{1.2cm}p{3.2cm}p{3.4cm}p{1.2cm}p{2.2cm}}
   \hline\hline
  &$D(2D_2^\prime)$as $c\bar{u} $ &&& $D(2D_2^\prime)$ as $c\bar{d}$\\
   \hline

   & Channels                            & Width    & $  $   &Channels                              &   Width    & $     $   \\
   \hline
   &$D^0\rho(770)^0$                     &  $1.8  $ & $  $   &$D^0\rho(770)^+$                      &   $3.7 $   & $  $   \\  
   &$D^+\rho(770)^-$                     &  $3.7  $ & $  $   &$D^+\rho(770)^0$                      &   $1.9 $   & $  $   \\  
   &$D^0\phi(1020) $                     &  $1.6  $ & $  $   &$D^+\phi(1020) $                      &   $1.5 $   & $  $   \\  
   &$D^0\omega  $                        &  $1.9  $ & $  $   &$D^+\omega  $                         &   $1.9 $   & $  $   \\  
   &$D_s^+K^*(892)^-$                    &  $1.2  $ & $  $   &$D_s^+K^*(892)^0$                     &   $1.2 $   & $  $   \\  

   &$D^0b_1(1235)^0$                     &  $0.2  $ & $  $   &$D^0b_1(1235)^+ $                     &   $0.3 $   & $  $   \\  
   &$D^+b_1(1235)^-$                     &  $0.3  $ & $  $   &$D^+b_1(1235)^0$                      &   $0.2 $   & $  $   \\  
   &$D^0a_1(1260)^0$                     &  $2.5  $ & $  $   &$D^0a_1(1260)^+$                      &   $5.0 $   & $  $   \\  
   &$D^+a_1(1260)^-$                     &  $4.7  $ & $  $   &$D^+a_1(1260)^0$                      &   $2.4 $   & $  $   \\  

   &$D^0a_2(1320)^0$                     &  $0.3  $ & $  $   &$D^0a_2(1320)^+$                      &   $0.7 $   & $  $   \\  
   &$D^+a_2(1320)^-$                     &  $0.7  $ & $  $   &$D^+a_2(1320)^0$                      &   $0.3 $   & $  $   \\  

   &$D^{*0}\pi^0$                        &  $6.4  $ & $  $   &$D^{*0}\pi^+$                         &   $12.7$   & $  $   \\  
   &$D^{*+}\pi^-$                        &  $12.6 $ & $  $   &$D^{*+}\pi^0$                         &   $6.3 $   & $  $   \\  
   &$D^{*0}\eta $                        &  $1.9  $ & $  $   &$D^{*+}\eta $                         &   $1.8 $   & $  $   \\  
   &$D^{*0}\eta^\prime $                 &  $0.2  $ & $  $   &$D^{*+}\eta^\prime $                  &   $0.2 $   & $  $   \\  
   &$D_s^{*+}K^-$                        &  $0.0  $ & $  $   &$D_s^{*+}K^0$                         &   $0.0 $   & $  $   \\  
   &$D_{s1}^*(2700)^+K^-$                &  $0.2  $ & $  $   &$D_{s1}^*(2700)^+K^0$                 &   $0.2 $   & $  $   \\  

   &$D^{*0}\rho(770)^0$                  &  $5.7  $ & $  $   &$D^{*0}\rho(770)^+$                   &   $11.4$   & $  $   \\  
   &$D^{*+}\rho(770)^-$                  &  $11.2 $ & $  $   &$D^{*+}\rho(770)^0$                   &   $5.6 $   & $  $   \\  
   &$D^{*0}\phi(1020) $                  &  $2.7  $ & $  $   &$D^{*+}\phi(1020) $                   &   $2.6 $   & $  $   \\  
   &$D^{*0}\omega $                      &  $5.6  $ & $  $   &$D^{*+}\omega  $                      &   $5.5 $   & $  $   \\  
   &$D_s^{*+}K^*(892)^-$                 &  $3.1  $ & $  $   &$D_s^{*+}K^*(892)^0$                  &   $3.0 $   & $  $   \\  


   &$D_0^*(2400)^0\pi^0$                 &  $0.4  $ & $  $   &$D_0^*(2400)^0\pi^+$                  &   $0.8 $   & $  $   \\  
   &$D_0^*(2400)^+\pi^-$                 &  $0.9  $ & $  $   &$D_0^*(2400)^+\pi^0$                  &   $0.5 $   & $  $   \\  
   &$D_0^*(2400)^0\eta $                 &  $0.3  $ & $  $   &$D_0^*(2400)^+\eta $                  &   $0.3 $   & $  $   \\  
   &$D_{s0}^*(2317)^{+}K^-$              &  $0.8  $ & $  $   &$D_{s0}^*(2317)^{+}K^0$               &   $0.8 $   & $  $   \\  
   &$D(2^3P_0)^0\pi^0$                   &  $0.0  $ & $  $   &$D(2^3P_0)^0\pi^+$                    &   $0.0 $   & $  $   \\  
   &$D(2^3P_0)^+\pi^-$                   &  $0.0  $ & $  $   &$D(2^3P_0)^+\pi^0$                    &   $0.0 $   & $  $   \\  

   &$D_0^*(2400)^0\rho(770)^0$           &  $0.6  $ & $  $   &$D_0^*(2400)^0\rho(770)^+$            &   $1.2 $   & $  $   \\  
   &$D_0^*(2400)^+\rho(770)^-$           &  $0.3  $ & $  $   &$D_0^*(2400)^+\rho(770)^0$            &   $0.1 $   & $  $   \\  
   &$D_0^*(2400)^0\omega $               &  $0.6  $ & $  $   &$D_0^*(2400)^+\omega $                &   $0.1 $   & $  $   \\  
   &$D_{s0}^*(2317)^{+}K^*(892)^-$       &  $0.0  $ & $  $   &$D_{s0}^*(2317)^{+}K^*(892)^0$        &   $0.0 $   & $  $   \\  

   &$D_1(2420)^0\pi^0$                   &  $1.3  $ & $  $   &$D_1(2420)^0\pi^+$                    &   $2.6 $   & $  $   \\  
   &$D_1(2420)^+\pi^-$                   &  $2.6  $ & $  $   &$D_1(2420)^+\pi^0$                    &   $1.3 $   & $  $   \\  
   &$D_1(2420)^0\eta$                    &  $0.8  $ & $  $   &$D_1(2420)^+\eta$                     &   $0.8 $   & $  $   \\  
   &$D_{s1}(2536)^+K^-$                  &  $0.4  $ & $  $   &$D_{s1}(2536)^+K^0$                   &   $0.4 $   & $  $   \\  
   &$D_1(2430)^0\pi^0$                   &  $1.1  $ & $  $   &$D_1(2430)^0\pi^+$                    &   $2.2 $   & $  $   \\  
   &$D_1(2430)^+\pi^-$                   &  $2.2  $ & $  $   &$D_1(2430)^+\pi^0$                    &   $1.1 $   & $  $   \\  
   &$D_1(2430)^0\eta$                    &  $0.6  $ & $  $   &$D_1(2430)^+\eta$                     &   $0.6 $   & $  $   \\  
   &$D_{s1}(2460)^{+}K^-$                &  $3.1  $ & $  $   &$D_{s1}(2460)^{+}K^0$                 &   $3.0 $   & $  $   \\  
   &$D(2P_1)^0\pi^0$                     &  $0.3  $ & $  $   &$D(2P_1)^0\pi^+$                      &   $0.6 $   & $ $   \\  
   &$D(2P_1)^+\pi^-$                     &  $0.6  $ & $  $   &$D(2P_1)^+\pi^0$                      &   $0.3 $   & $ $   \\  
   &$D(2P_1^\prime)^0\pi^0$              &  $0.0  $ & $  $   &$D(2P_1^\prime)^0\pi^+$               &   $0.0 $   & $ $   \\  
   &$D(2P_1^\prime)^+\pi^-$              &  $0.0  $ & $  $   &$D(2P_1^\prime)^+\pi^0$               &   $0.0 $   & $ $   \\  

   &$D_1(2420)^0\rho(770)^0$             &  $0.8  $ & $  $   &$D_1(2420)^0\rho(770)^+$              &   $1.5 $   & $  $   \\  
   &$D_1(2420)^+\rho(770)^-$             &  $1.5  $ & $  $   &$D_1(2420)^+\rho(770)^0$              &   $0.8 $   & $  $   \\  
   &$D_1(2420)^0\omega$                  &  $0.8  $ & $  $   &$D_1(2420)^+\omega$                   &   $0.8 $   & $  $   \\  
   &$D_1(2430)^0\rho(770)^0$             &  $0.3  $ & $  $   &$D_1(2430)^0\rho(770)^+$              &   $0.6 $   & $  $   \\  
   &$D_1(2430)^+\rho(770)^-$             &  $0.6  $ & $  $   &$D_1(2430)^+\rho(770)^0$              &   $0.3 $   & $  $   \\  
   &$D_1(2430)^0\omega$                  &  $0.2  $ & $  $   &$D_1(2430)^+\omega$                   &   $0.2 $   & $  $   \\  

   &$D_2^*(2460)^0\pi^0$                 &  $11.9 $ & $  $   &$D_2^*(2460)^0\pi^+$                  &   $23.8$   & $  $   \\  
   &$D_2^*(2460)^+\pi^-$                 &  $23.8 $ & $  $   &$D_2^*(2460)^+\pi^0$                  &   $11.9$   & $  $   \\  
   &$D_2^*(2460)^0\eta $                 &  $2.7  $ & $  $   &$D_2^*(2460)^+\eta $                  &   $2.7 $   & $  $   \\  
   &$D_{s2}^*(2573)^+K^-$                &  $12.7 $ & $  $   &$D_{s2}^*(2573)^+K^0$                 &   $13.0$   & $  $   \\  
   &$D(2^3P_2)^0\pi^0$                   &  $6.6  $ & $  $   &$D(2^3P_2)^0\pi^+$                    &   $13.4$   & $  $   \\  
   &$D(2^3P_2)^+\pi^-$                   &  $13.4 $ & $  $   &$D(2^3P_2)^+\pi^0$                    &   $6.6 $   & $  $   \\  


   & $\Gamma^\prime$                    &  $173.2$ & $  $   & $\Gamma^\prime$                       &   $173.1$  & $  $   \\

         \hline\hline
\end{tabular}
\label{table5}
\end{table*}

\begin{table*}[t]
\caption{Hadronic decay widths of $D(2D_2^\prime)$ (in MeV).}
\begin{tabular}{p{0cm}p{3.4cm}p{1.2cm}p{3.2cm}p{3.4cm}p{1.2cm}p{2.2cm}}
   \hline\hline
  &$D(2D_2^\prime)$as $c\bar{u} $ &&& $D(2D_2^\prime)$ as $c\bar{d}$\\
   \hline

   & Channels                            & Width    & $  $   &Channels                              &   Width    & $     $   \\
   \hline
   &$D^*(2600)^0(2^3S_1)\pi^0$             &  $3.6  $ & $  $   &$D^*(2600)^0(2^3S_1)\pi^+$              &   $7.3 $   & $  $   \\  
   &$D^*(2600)^+(2^3S_1)\pi^-$             &  $6.9  $ & $  $   &$D^*(2600)^+(2^3S_1)\pi^0$              &   $3.4 $   & $  $   \\  
   &$D^*(2600)^0(1^3D_1)\pi^0$             &  $0.1  $ & $  $   &$D^*(2600)^0(1^3D_1)\pi^+$              &   $0.1 $   & $  $   \\  
   &$D^*(2600)^+(1^3D_1)\pi^-$             &  $0.1  $ & $  $   &$D^*(2600)^+(1^3D_1)\pi^0$              &   $0.1 $   & $  $   \\  
   &$                      $             &  $     $ & $     $   &$                      $              &   $    $   & $     $   \\

   &$D(2750)^0(1^1D_2)\pi^0$             &  $1.1  $ & $  $   &$D(2750)^0(1^1D_2)\pi^+$              &   $2.3 $   & $  $   \\  
   &$D(2750)^0(1^3D_2)\pi^0$             &  $0.9  $ & $  $   &$D(2750)^0(1^3D_2)\pi^+$              &   $1.8 $   & $  $   \\  
   &$                      $             &  $     $ & $     $   &$                      $              &   $    $   & $     $   \\


   &$D_1^*(2760)^0(1^3D_1)\pi^0$         &  $0.0  $ & $  $   &$D_1^*(2760)^0(1^3D_1)\pi^+$          &   $0.0 $   & $ $   \\  
   &$D_1^*(2760)^+(1^3D_1)\pi^-$         &  $0.0  $ & $  $   &$D_1^*(2760)^+(1^3D_1)\pi^0$          &   $0.0 $   & $ $   \\  
   &$D_3^*(2760)^0(1^3D_3)\pi^0$         &  $14.4 $ & $  $   &$D_3^*(2760)^0(1^3D_3)\pi^+$          &   $28.3$   & $ $   \\  
   &$D_3^*(2760)^+(1^3D_3)\pi^-$         &  $28.8 $ & $  $   &$D_3^*(2760)^+(1^3D_3)\pi^0$          &   $14.2$   & $ $   \\  

         \hline\hline
\end{tabular}
\label{table5}
\end{table*}

\subsection{Mixing of $1P$ and $2D$}

\par As well known, there is mixing between $1^1P_1$ state and $1^3P_1$ state. $D_1(2420)$ and $D_1(2430)$ are believed two mixing states, and the mixing follows~\cite{gm}
\begin{eqnarray*}
\left(
\begin{array}{c}
|1^+,j^p=\frac{1}{2}^+\rangle\\
  \\
|1^+,j^p=\frac{3}{2}^+\rangle
\end{array}
\right)=\left(\begin{array}{c}
cos\theta  \quad  sin\theta\\
\displaystyle
-sin\theta  \quad cos\theta   \\
\end{array}
\right)\left(
\begin{array}{c}
|1^1P_1\rangle\\
    \displaystyle
|1^3P_1 \rangle
\end{array}
\right).
\end{eqnarray*}
Similar mixing mechanism happens to $2^1P_1$ state and $2^3P_1$ state, where the mixing angle $\theta_{1P}=-25.68^\circ$, $\theta_{2P}=-29.39^\circ$.
\par
$2^1D_2$ state may mix with $2^3D_2$ state. To take into account this mixing, $2D_2$ and $2D^\prime_2$ are assumed as two mixing states as follows~\cite{gm},
\begin{eqnarray*}
\label{eqn:hqlmixing}
|J= L, j_q=L + \frac{1}{2} \rangle & = & \sqrt{{J+1}\over{2J+1}} | J=L, S=0 \rangle \nonumber \\
			& &  + \sqrt{\frac{J}{2J+1}} | J=L, S=1 \rangle \nonumber \\
|J= L, j_q=L - \frac{1}{2} \rangle & = & - \sqrt{{J}\over{2J+1}} | J=L, S=0 \rangle \nonumber \\
			& &  + \sqrt{{J+1}\over{2J+1}} | J=L, S=1 \rangle
\end{eqnarray*}
where $j_q=L-\frac{1}{2}$ state (mainly spin triplet) corresponds to the primed state $2D^\prime_2$ and $j_q=L+\frac{1}{2}$ (mainly spin singlet) corresponds to the unprimed state $2D_2$. These mixing schemes are employed in our calculation.

\section{Analysis of $D(2550)$, $D^*(2600)$, $D(2750)$, $D^*_1(2760)$ and $D^*_3(2760)$}

\subsection{$D(2550)$}
\par $D(2550)$ is very possibly $2S$ $^1S_0$ with $J^P=0^-$. Decay channel $D(2550)\pi$ is expected to be observed in hadronic decays from $D(2^3D_1)$ and $D(2^3D_3)$. In particular, we find the following ratios:
$$ \frac{\Gamma_{D(2^3D_1)^0 \to D(2550)^0\pi^0}}{\Gamma_{D(2^3D_1)^0 \to D^{*0}\pi^0}}=0.71, \frac{\Gamma_{D(2^3D_1)^0 \to D(2550)^+\pi^-}}{\Gamma_{D(2^3D_1)^0 \to D^{*+}\pi^-}}=0.76,$$

$$ \frac{\Gamma_{D(2^3D_3)^0 \to D(2550)^0\pi^0}}{\Gamma_{D(2^3D_3)^0 \to D^{*0}\pi^0}}=6.00. \frac{\Gamma_{D(2^3D_3)^0 \to D(2550)^+\pi^-}}{\Gamma_{D(2^3D_3)^0 \to D^{*+}\pi^-}}=5.55.$$

\subsection{$D^*(2600)$}
\par
$D^*(2600)$ is very possibly a $2^3S_1$ state~\cite{Benitez,wang,colangelo,chen2,gm}, it is also suggested as an admixture of $2^3S_1$ and $1^3D_1$ with $J^P=1^-$~\cite{liu3,zhong}.
\par
If $D^*(2600)$ is a $2^3S_1$ state, it is expected to be observed in hadronic decays from all four excited $2D$ $D$ states. We obtain the following ratios:
$$ \frac{\Gamma_{D(2^3D_1)^0 \to D^*(2600)^0\pi^0}}{\Gamma_{D(2^3D_1)^0 \to D^{*0}\pi^0}}=0.76, \frac{\Gamma_{D(2^3D_1)^0 \to D^*(2600)^+\pi^-}}{\Gamma_{D(2^3D_1)^0 \to D^{*+}\pi^-}}=0.76,$$

$$ \frac{\Gamma_{D(2^3D_3)^0 \to D^*(2600)^0\pi^0}}{\Gamma_{D(2^3D_3)^0 \to D^{*0}\pi^0}}=5.00, \frac{\Gamma_{D(2^3D_3)^0 \to D^*(2600)^+\pi^-}}{\Gamma_{D(2^3D_3)^+ \to D^{*+}\pi^-}}=4.91,$$

$$ \frac{\Gamma_{D(2D_2)^0 \to D^*(2600)^0\pi^0}}{\Gamma_{D(2D_2)^0 \to D^{*0}\pi^0}}=1.77, \frac{\Gamma_{D(2D_2)^0 \to D^*(2600)^+\pi^-}}{\Gamma_{D(2D_2)^+ \to D^{*+}\pi^-}}=1.80,$$

$$ \frac{\Gamma_{D(2D^\prime_2)^0 \to D^*(2600)^0\pi^0}}{\Gamma_{D(2D^\prime_2)^0 \to D^{*0}\pi^0}}=0.82, \frac{\Gamma_{D(2D^\prime_2)^0 \to D^*(2600)^+\pi^-}}{\Gamma_{D(2D^\prime_2)^0 \to D^{*+}\pi^-}}=0.55.$$
If $D^*(2600)$ is a $1^3D_1$ state, the hadronic decay width from $D(2^3D_1)$ to $D^*(2600)\pi$ channel vanishes, and the hadronic decay width from other three excited $D(2D)$ resonances to $D^*(2600)\pi$ channel are very small. We obtain the following ratios:
$$ \frac{\Gamma_{D(2^3D_3)^0 \to D^*(2600)^0\pi^0}}{\Gamma_{D(2^3D_3)^0 \to D^{*0}\pi^0}}=0.20, \frac{\Gamma_{D(2^3D_3)^0 \to D^*(2600)^+\pi^-}}{\Gamma_{D(2^3D_3)^0 \to D^{*+}\pi^-}}\approx0.00,$$

$$ \frac{\Gamma_{D(2D_2)^0 \to D^*(2600)^0\pi^0}}{\Gamma_{D(2D_2)^0 \to D^{*0}\pi^0}}=0.09, \frac{\Gamma_{D(2D_2)^0 \to D^*(2600)^+\pi^-}}{\Gamma_{D(2D_2)^0 \to D^{*+}\pi^-}}=0.09,$$

$$ \frac{\Gamma_{D(2D^\prime_2)^0 \to D^*(2600)^0\pi^0}}{\Gamma_{D(2D^\prime_2)^0 \to D^{*0}\pi^0}}=0.02, \frac{\Gamma_{D(2D^\prime_2)^0 \to D^*(2600)^+\pi^-}}{\Gamma_{D(2D^\prime_2)^0 \to D^{*+}\pi^-}}=0.01.$$

If $D^*(2600)$ is an admixture of $2^3S_1$ and $1^3D_1$ with $J^P=1^-$, the case is a little more complicated, where the numerical ratios depend on detail of the mixing.

Obviously, in different assignments, the ratios are different. In fact, these ratios may provide another way to identify these resonance in the future experiments.
\subsection{$D(2750)$}
\par
$D(2750)$ is possibly a $1^1D_2$ or a $1^3D_2$ state, it is also possibly an admixture of them~\cite{zhong,wang,colangelo,chen2,gm}. In any assignment, $D(2750)$ can be observed in $D(2750)\pi$ channel from hadronic decays of the four excited $2D$ states. If $D(2750)$ is a $1^1D_2$, we obtain the following ratios:
$$ \frac{\Gamma_{D(2^3D_1)^0 \to D(2750)^0\pi^0}}{\Gamma_{D(2^3D_1)^0 \to D^{*0}\pi^0}}=7.29,$$

$$ \frac{\Gamma_{D(2^3D_3)^0 \to D(2750)^0\pi^0}}{\Gamma_{D(2^3D_3)^0 \to D^{*0}\pi^0}}=1.20,$$

$$ \frac{\Gamma_{D(2D_2)^0 \to D(2750)^0\pi^0}}{\Gamma_{D(2D_2)^0 \to D^{*0}\pi^0}}=0.32,$$

$$ \frac{\Gamma_{D(2D^\prime_2)^0 \to D(2750)^0\pi^0}}{\Gamma_{D(2D^\prime_2)^0 \to D^{*0}\pi^0}}=0.17.$$

If $D(2750)$ is a $1^3D_2$, we obtain the following ratios:
$$ \frac{\Gamma_{D(2^3D_1)^0 \to D(2750)^0\pi^0}}{\Gamma_{D(2^3D_1)^0 \to D^{*0}\pi^0}}=5.06,$$

$$ \frac{\Gamma_{D(2^3D_3)^0 \to D(2750)^0\pi^0}}{\Gamma_{D(2^3D_3)^0 \to D^{*0}\pi^0}}=0.20,$$

$$ \frac{\Gamma_{D(2D_2)^0 \to D(2750)^0\pi^0}}{\Gamma_{D(2D_2)^0 \to D^{*0}\pi^0}}=0.36,$$

$$ \frac{\Gamma_{D(2D^\prime_2)^0 \to D(2750)^0\pi}}{\Gamma_{D(2D^\prime_2)^0 \to D^{*0}\pi^0}}=0.14.$$
Similar to the case of $D^*(2600)$, if $D(2750)$ is an admixture, the numerical results will depend on the detail of the mixing.

\subsection{$D^*_1(2760)$ and $D^*_3(2760)$}
$D^*_1(2760)$ and $D^*_3(2760)$ are regarded as $1^3D_1$ predominantly and $1^3D_3$~\cite{chen3,gm}, respectively. $D^*_1(2760)$ can be observed in $D(2D_2)$, while cannot be observed from the hadronic decays of $D(2^3D_1)$, $D(2^3D_3)$, or $D(2D^\prime_2)$. We have the ratio:
$$ \frac{\Gamma_{D(2D_2)^0 \to D^*_1(2760)^0\pi^0}}{\Gamma_{D(2D_2)^0 \to D^{*0}\pi^0}}=0.14, \frac{\Gamma_{D(2D_2)^0 \to D^*_1(2760)^+\pi^-}}{\Gamma_{D(2D_2)^0 \to D^{*+}\pi^-}}=0.13.$$

$D^*_3(2760)$ can be observed from the hadronic decays of the four excited $D(2^3D_1)$, $D(2^3D_3)$, $D(2D_2)$ and $D(2D^\prime_2)$. We have the ratios:
$$ \frac{\Gamma_{D(2^3D_1)^0 \to D^*_3(2760)^0\pi^0}}{\Gamma_{D(2^3D_1)^0 \to D^{*0}\pi^0}}=0.18, \frac{\Gamma_{D(2^3D_1)^0 \to D^*_3(2760)^+\pi^-}}{\Gamma_{D(2^3D_1)^0 \to D^{*+}\pi^-}}=0.18,$$

$$ \frac{\Gamma_{D(2^3D_3)^0 \to D^*_3(2760)^0\pi^0}}{\Gamma_{D(2^3D_3)^0 \to D^{*0}\pi^0}}=1.80, \frac{\Gamma_{D(2^3D_3)^0 \to D^*_3(2760)^+\pi^-}}{\Gamma_{D(2^3D_3)^0 \to D^{*+}\pi^-}}=1.64,$$

$$ \frac{\Gamma_{D(2D_2)^0 \to D^*_3(2760)^0\pi^0}}{\Gamma_{D(2D_2)^0 \to D^{*0}\pi^0}}=0.09, \frac{\Gamma_{D(2D_2)^0 \to D^*_3(2760)^+\pi^-}}{\Gamma_{D(2D_2)^0 \to D^{*+}\pi^-}}=0.07,$$

$$ \frac{\Gamma_{D(2D^\prime_2)^0 \to D^*_3(2760)^0\pi^0}}{\Gamma_{D(2D^\prime_2)^0 \to D^{*0}\pi^0}}=2.25, \frac{\Gamma_{D(2D^\prime_2)^0 \to D^*_3(2760)^+\pi^-}}{\Gamma_{D(2D^\prime_2)^0 \to D^{*+}\pi^-}}=2.29.$$

\section{CONCLUSIONS AND DISCUSSIONS}
\par In this paper, the radially excited $2D$ $D(2^3D_1)$, $D(2^3D_3)$, $D(2D_2)$ and $D(2D^\prime_2)$ have been studied in the $^3P_0$ model. Uncertainties resulting from the parameters in $^3P_0$ model have not been explored in our study. The hadronic decay widths of all possible $OZI$-allowed channels of these resonances have been calculated, and some relevant ratios of branching fraction have been given. For lack of experimental data, the detail of mixing has not been explored.

The dominant decays of $D(2^3D_1)$ are $D(2^3D_1)\to D^*\rho(770)$, $D(2^3D_1)\to D_1(2420)\pi$, $D(2^3D_1)\to D\pi(1300)$, $D(2^3D_1)\to D_1(2420)\rho(770)$, and $D(2^3D_1)\to D^*_2(2460)\pi$. The dominant decays of $D(2^3D_3)$ are $D(2^3D_3)\to D^*_2(2460)\pi$, $D(2^3D_3)\to D^*\rho(770)$, $D(2^3D_3)\to D_1(2420)\pi$, and $D(2^3D_3)\to D(2550)\pi$. The dominant decays of $D(2D_2)$ are $D(2D_2)\to D_1(2420)\rho(770)$, $D(2D_2)\to Da_2(1320)$, $D(2D_2)\to D^*_2(2460)\pi$, and $D(2D_2)\to D^*\rho(770)$. The dominant decays of $D(2D^\prime_2)$ are $D(2D_2^\prime)\to D^*_2(2460)\pi$, $D(2D_2^\prime)\to D(2^3P_2)\pi$, $D(2D_2^\prime)\to D^*\pi$, and $D(2D_2^\prime)\to D^*\rho(770)$.
\par
$D(2550)$, $D^*(2600)$, $D(2750)$, $D^*_1(2760)$ and $D^*_3(2760)$ can be produced in hadronic decays of $2D$ $D(2^3D_1)$, $D(2^3D_3)$, $D(2D_2)$ and $D(2D^\prime_2)$. The hadronic decay widths and some relevant ratios from the radially excited $2D$ $D$ resonances to these resonances are predicted. As more and more highly excited charmed resonances have been observed, the radially excited $2D$ $D$ resonances are expected to be observed in the future B factories and LHCb. The hadronic production of $D(2550)$, $D^*(2600)$, $D(2750)$, $D^*_1(2760)$ and $D^*_3(2760)$ from radially excited $2D$ $D$ resonances provides another way to identify these resonances.
\par
In addition to $D^*_1(2760)$ and $D^*_3(2760)$ with unnatural parity, the natural parity $D(2750)$ with mass close to them has been observed. As suggested in Ref.~\cite{chen2}, a natural parity $D_{sJ}(2850)$ close to $D^*_{sJ}(2860)$ may exist. However, no charmed strange resonance located around $2860$ MeV with natural parity has been observed. Therefore, $D(2750)$ has to be explored in detail by experiment, and charmed strange sector around $2860$ MeV has to be studied more. Of course, $D(2550)$, $D^*(2600)$, $D(2750)$, $D^*_1(2760)$ and $D^*_3(2760)$ may be produced in hadronic decays of other highly excited resonances, which requires more exploration. 

\begin{acknowledgments}
This work is supported by National Natural Science Foundation of China under the grants:11075102 and 11475111. It is also supported by the Innovation Program of Shanghai Municipal Education Commission under the grant No.13ZZ066.
\end{acknowledgments}

\end{document}